\documentclass{article}
\usepackage[utf8]{inputenc}
\usepackage{fullpage}
\usepackage{setspace}
\usepackage{amsmath}              
\usepackage{amssymb}
\usepackage{bm}                   
\usepackage{mathtools}            
\usepackage{color, soul}                    
\usepackage[usenames,dvipsnames]{xcolor}    
\usepackage{url, hyperref} 
\usepackage{natbib}
\usepackage{float}
\usepackage{scalerel,stackengine}  
\usepackage{booktabs} 
\usepackage{graphicx}
\usepackage{authblk}
\newtheorem{thm}{Theorem}[section]

\newcommand*{\QEDB}{\hfill\ensuremath{\square}}%
\usepackage{verbatim}
\usepackage{placeins}
\usepackage{algorithm, algpseudocode}
\usepackage{comment}

\stackMath
\newcommand\reallywidehat[1]{%
\savestack{\tmpbox}{\stretchto{%
  \scaleto{%
    \scalerel*[\widthof{\ensuremath{#1}}]{\kern-.6pt\bigwedge\kern-.6pt}%
    {\rule[-\textheight/2]{1ex}{\textheight}}
  }{\textheight}%
}{0.5ex}}%
\stackon[1pt]{#1}{\tmpbox}%
}

\DeclareMathOperator*{\argmin}{arg\,min}
\DeclareMathOperator{\E}{\textnormal{\mbox{E}}}
\DeclareMathOperator{\PP}{\textnormal{\mbox{P}}}

\algnewcommand{\Initialize}[1]{%
  \Statex \textbf{Initialize:}
  \Statex \hspace*{\algorithmicindent}\parbox[t]{0.95\linewidth}{\raggedright #1}
}
\algnewcommand{\Result}[1]{%
  \Statex \textbf{Result:}
  \Statex \hspace*{\algorithmicindent}\parbox[t]{0.95\linewidth}{\raggedright #1}
}

\title{Causal Interaction Trees: Tree-Based Subgroup Identification for Observational Data}
\author[1]{Jiabei Yang}
\author[2-3]{Issa J. Dahabreh}
\author[1]{Jon A. Steingrimsson}
\affil[1]{Department of Biostatistics, School of Public Health, Brown University}
\affil[2]{Center for Evidence Synthesis in Health, School of Public Health, Brown University}
\affil[3]{Departments of Health Services, Policy \& Practice and Epidemiology, School of Public Health, Brown University}
\date{}   

\begin{document}

\maketitle



\textbf{Abstract:} We propose Causal Interaction Trees for identifying subgroups of participants that have enhanced treatment effects using observational data. We extend the Classification and Regression Tree algorithm by using splitting criteria that focus on maximizing between-group treatment effect heterogeneity based on subgroup-specific treatment effect estimators to dictate decision-making in the algorithm. We derive properties of three subgroup-specific treatment effect estimators that account for the observational nature of the data -- inverse probability weighting, g-formula and doubly robust estimators. We study the performance of the proposed algorithms using simulations and implement the algorithms in an observational study that evaluates the effectiveness of right heart catheterization on critically ill patients.
\vspace{0.3cm}



\section{Introduction}

Subgroup identification in randomized trials aims to identify subsets of participants that have enhanced treatment effects, allowing more targeted treatment recommendations. This is typically done by performing subgroup analyses or by exploring a few treatment-covariate interactions using generalized linear models \citep{dahabreh2016using, dahabreh2017heterogeneity}. A non-parametric data driven alternative for exploring treatment-covariate interactions is to use extensions of the Classification and Regression Tree algorithm \citep{breiman1984classification} appropriate for subgroup identification \citep{su2009subgroup, foster2011subgroup, seibold2016model, steingrimsson2019subgroup}. Tree-based methods recursively partition the covariate space using splitting criteria until some pre-determined stopping criteria are met, creating a large, potentially overfit, tree that can be used as a prediction model. To reduce overfitting, a subtree of the large tree is selected using pruning criteria. Tree-based methods are appealing for subgroup identification because they can identify treatment-covariate interactions without having to pre-specify the interactions to include in the model. An early example of a tree-based algorithm for subgroup identification in randomized trials is the interaction tree algorithm \citep{su2009subgroup}. The algorithm makes splitting decisions by contrasting treatment effect estimates between groups, where the treatment effects are estimated by differences in outcomes between the treatment arms. 
\cite{lipkovich2017tutorial} provides a review of data-driven subgroup identification methods for randomized trials.

When the treatment is not randomly assigned to participants, confounding of the treatment-outcome relationship complicates subgroup identification. Less work has focused on the use of tree-based methods for subgroup identification with observational data. \cite{su2012facilitating}, \cite{kang2012tree}, and \cite{kang2014causal} proposed a likelihood-based splitting statistic with AIC-type pruning criteria, which requires specifying the distribution of the outcome conditional on the covariates. \cite{athey2016recursive} proposed the Causal Tree algorithm where splitting decisions are based on minimizing an estimator of the mean squared error of the subgroup-specific treatment effect. \cite{wager2018estimation} used the Causal Tree algorithm to build a random forest algorithm, an ensemble method that averages multiple fully grown trees. Finally, \cite{powers2018some} proposed ensemble methods to control confounding using inverse probability of treatment weighting. Ensemble methods focus on a different objective than single trees because they create black-box individualized prediction models, rather than clinically interpretable subgroups. 

In this paper, we develop three new tree-based algorithms, the Causal Interaction Tree (CIT) algorithms, for subgroup identification with observational data. All are generalizations of the interaction tree algorithm that utilize inverse probability weighting, g-formula, or doubly robust estimators of subgroup-specific treatment effects for decision-making during tree construction. Here, doubly robust refers to estimators that are consistent when either the outcome model or the propensity score model required for implementation are correctly specified. In contrast, consistency of the inverse probability weighting estimator requires a correctly specified propensity score model and consistency of the g-formula estimators requires a correctly specified outcome model.

In Section \ref{sec: IT}, we define the Generalized Interaction Tree (GIT) algorithm that includes, as special cases, the original interaction tree algorithm of \cite{su2009subgroup} and the covariate adjusted interaction tree algorithm of \cite{steingrimsson2019subgroup} for randomized trials, and the three Causal Interaction Tree algorithms for observational studies. In Section \ref{sec: estimators}, we derive the properties of the three subgroup-specific treatment effect estimators that dictate decision-making in the Causal Interaction Tree algorithms. The algorithms are implemented by modifying the \texttt{rpart} package, the most popular implementation of tree-based methods in \texttt{R}, to accommodate the different node-specific treatment effect estimators. We evaluate the performance of the subgroup identification methods through simulations and analyses of data from an observational study that evaluated the effectiveness of right heart catheterization on critically ill patients. Results from the simulations and the data analyses are presented in Sections \ref{sec: simulations} and \ref{sec:data-analysis}, respectively. The Supplementary Web Appendix contains proofs, additional simulation results, and additional details about the data analysis.

\section{Generalized Interaction Tree Algorithm} \label{sec: IT}

Let $Y$ be an outcome measured at the end of the study (binary, continuous, or count); $A$ be an indicator for exposure to treatment which equals 1 when the participant is exposed and equals 0 when not exposed; $\bm{X}$ be a vector of pre-exposure covariates taking values in $\mathcal{X}$. A set $w$ is called a subgroup if $w$ is a subset of $\mathcal{X}$. The data collected is assumed to consist of $n$ i.i.d.~observations of $(\bm{X}, A, Y)$. Let $n(w) = \sum_{i=1}^n I(\bm{X}_i \in w)$ be the number of observations in subgroup $w$.
Let $Y^{a}$ be the potential outcome under intervention to set treatment to $a$, $a \in \{0,1\}$ \citep{rubin1974estimating, robins2000causal}. The average treatment effect for subgroup $w$ is defined as $\E[Y^{1}| \bm{X}\in w] - \E[Y^{0}| \bm{X}\in w] =\mu_1(w) - \mu_0(w)$, where  $\mu_a(w) = \E[Y^{a}| \bm{X}\in w]$ for $a \in \{0,1\}$. We are interested in finding a set of subgroups $w_1, \ldots, w_{\hat K}$ that stratify observations into a finite set of mutually exclusive groups based on their treatment effects and the union of the subgroups is exhaustive of $\mathcal X$. The number of identified subgroups, $\hat K$, is estimated from the data (see Steps 2 and 3 of the Generalized Interaction Tree algorithm defined below). Implementation of the algorithm relies on estimating $\mu_a(w)$ for $a \in \{0,1\}$ and we use $\hat \mu_a(w)$ to denote a general estimator of $\mu_a(w)$. In Section \ref{sec: estimators}, we describe three estimators for $\mu_a(w)$ that can be used with observational data.

\cite{su2009subgroup} proposed the interaction tree algorithm for use in randomized trials. We will now describe the Generalized Interaction Tree algorithm that will serve as the basis for extensions to observational studies. The Generalized Interaction Tree algorithm includes, as special cases, the original interaction tree algorithm of \cite{su2009subgroup} and the covariate adjusted interaction tree algorithm of \cite{steingrimsson2019subgroup}.

We summarize the Generalized Interaction Tree algorithm using pseudocode in Algorithm \ref{alg: git}. In detail, the algorithm consists of the following 3 steps:

\begin{enumerate}
    \item \textit{Creating a maximum sized tree:} At the beginning of the tree building process, all observations are in a single node, referred to as the root node. Define $X^{(j)}$ as the $j$-th component of the covariate vector $\bm{X}$, and let $c\in \mathbb{R}$. The pair $(X^{(j)},c)$ splits the covariate space into two groups $l = \{X^{(j)} < c\}$ and $r = \{X^{(j)} \geq c\}$, with corresponding treatment effects $\hat{\mu}_{1}(l) - \hat{\mu}_{0}(l)$ and $\hat{\mu}_{1}(r) - \hat{\mu}_{0}(r)$, respectively. We define the splitting statistic corresponding to this split as
    \begin{align} \label{eq: test_stat}
      \left(\frac{(\hat \mu_{1}(l) - \hat \mu_{0}(l)) - (\hat \mu_{1}(r) - \hat \mu_{0}(r))}{\sqrt{\reallywidehat{\text{Var}}\left[ (\hat \mu_{1}(l) - \hat \mu_{0}(l)) - (\hat \mu_{1}(r) - \hat \mu_{0}(r)) \right]}} \right)^2
    \end{align}
   The splitting statistic \eqref{eq: test_stat} measures a standardized difference between the treatment effect in the two groups. When the true treatment effect is identical in the two subgroups, the splitting statistic defined in expression (\ref{eq: test_stat}) converges to a $\chi^2$-distribution with 1 degree of freedom. The node that is being considered for splitting is referred to as the parent node and the two nodes that the data is split into are referred to as the child nodes.
    
    To split the node into two subgroups, the Generalized Interaction Tree algorithm cycles through all permissible $(X^{(j)}, c)$ pairs, and selects the combination that gives the largest splitting statistic in expression \eqref{eq: test_stat}. This process is iterated within each new subgroup until some pre-determined criteria are met. This procedure results in a large initial tree, $\hat{\psi}_{\max}$. For a categorical or an ordinal covariate, the algorithm will search through all possible combinations of levels for a categorical $X^{(j)}$, and all possible splits that preserve the ordering for an ordinal $X^{(j)}$.
    

    \item \textit{Pruning:} The pruning step creates a sequence of subtrees of $\hat{\psi}_{\max}$ that are candidates for being the final tree, reducing the computational complexity of the model selection process. This step is a modification of the original Classification and Regression Tree pruning algorithm \citep{breiman1984classification} that was adapted to the interaction tree setting in \cite{su2009subgroup}.
    
    In a given tree, nodes that are split are referred to as internal nodes; nodes that are not split are referred to as terminal nodes. Let a penalization parameter $\lambda$ be given and define the split complexity for a tree $\psi$ as
\begin{align} \label{eq: split_complex}
    G^{(\lambda)}(\psi) = \sum_{i \in I_\psi} G_i(\psi) - \lambda |I_{\psi}|.
\end{align}
Here, $G_i(\psi)$ is the value of the splitting statistic defined in expression \eqref{eq: test_stat} for internal node $i$ in tree $\psi$; $I_\psi$ is the set of internal nodes of $\psi$; and $|I_\psi|$ is the number of internal nodes. 

Weakest link pruning creates a finite sequence of subtrees of $\hat{\psi}_{\max}$ by sequentially dropping the branch of the tree that has the smallest split complexity. More formally, it is defined using the following three steps:

\begin{enumerate}
\item Set $\hat{\psi}_0 = \hat{\psi}_{\max}$ and $m = 0$.
\item Define $g(h) = \sum_{i \in I_{\psi^*_h}} G_i(\psi^*_h)/|I_{\psi^*_h}|$ if $h \in I_{\psi_k}$ and $g(h)=+\infty$ otherwise. Here, $\psi^*_h$ is the subtree consisting of node $h$ and all descendants of node $h$. The weakest link, defined in terms of split complexity, of the tree $\hat{\psi}_m$ is the node $h' = \argmin_{h \in I_{\psi_k}} g(h)$. Define $\hat{\psi}_{m+1}$ as the subtree of $\hat{\psi}_m$ with all the descendants of $h'$
removed. Set $m = m+1$.
\item Repeat Step (b) until $\hat{\psi}_{m+1}$ consists only of the root node.
\end{enumerate}
Running the weakest link pruning algorithm results in a sequence of trees  $\hat{\psi}_0 = \hat{\psi}_{\max}, \hat{\psi}_{1}, \ldots, \hat{\psi}_{\hat{M}}$.

\item \textit{Final Tree Selection:} The last step is to select a final tree from the sequence of candidate trees generated during the pruning step. To select the final tree, the dataset is split into an initial tree building dataset and a validation dataset. The initial tree building dataset is used to build the maximum sized tree and create the sequence of candidate trees (Steps one and two of the Generalized Interaction Tree algorithm). For a candidate tree $\hat{\psi}_m$, $m \in \{0, \dots, \hat{M}\}$, the split complexity (defined in equation \eqref{eq: split_complex}) is calculated using the validation set by sending each observation in the validation set down the candidate tree to calculate the splitting statistics defined in expression \eqref{eq: test_stat} for each internal node. The final tree is selected as the one that maximizes the validation split complexity for a fixed penalization parameter $\lambda$. A common criterion for selecting the penalization parameter is some quantile of the asymptotic distribution of the splitting statistic \eqref{eq: test_stat} when the treatment effect is identical in the two subgroups $l$ and $r$.
\end{enumerate}

The Generalized Interaction Tree algorithm partitions the covariate space into $\hat K$ terminal nodes $w_1, \ldots, w_{\hat K}$. The final tree based treatment effect estimator for terminal node $k$ is given by $\hat \mu_1(w_k) - \hat \mu_0(w_k)$.

\begin{algorithm}[htbp]
\setstretch{1.35}
\begin{algorithmic}[1]
\vspace{0.5mm}
\Result{A tree which partitions the covariate space into a set of mutually exclusive subgroups $w_1, \ldots, w_{\hat K}$.\\ Final tree based treatment effect estimator for each subgroup $w_{k}$ is $\hat \mu_1(w_k) - \hat \mu_0(w_k)$.}
\vspace{0.1mm}
\Initialize{Split the data into an initial tree building dataset and a validation dataset.}
\State Create a maximum sized tree $\hat{\psi}_{\max}$ using the initial tree building dataset:
\begin{enumerate}
    \item[(a)] Define the root node of tree $\hat{\psi}_{\max}$ as consisting of all observations in the initial tree building dataset. Set the root node as the node of interest.
    
    \item[(b)] In the node of interest, identify all permissible $(X^{(j)}, c)$ pairs that split the covariate space into two groups $l = \{X^{(j)} < c\}$ and $r = \{X^{(j)} \geq c\}$. 
    
    \item[(c)] Consider all such splits in (b) and divide the node of interest into two mutually exclusive subgroups using the split that gives the largest splitting statistic (as defined by expression \eqref{eq: test_stat}). 
    
    \item[(d)] Check pre-determined stopping criteria. If met, denote the current tree as $\hat{\psi}_{\max}$ and move to step 2 of the algorithm; otherwise, on every node that has not met the stopping criteria (and is not already split into child nodes), repeat Step 1(b)-1(d).
    
\end{enumerate}
\State Prune $\hat{\psi}_{\max}$ to create a sequence of candidate trees, $\hat{\psi}_0, \hat{\psi}_1,\dots, \hat{\psi}_{\hat{M}}$:

\begin{enumerate}
\item[(a)] Set $m = 0$ and $\hat{\psi}_0 = \hat{\psi}_{\max}$.

\item[(b)] Define $\hat{\psi}_{m+1}$ as the subtree of $\hat{\psi}_m$ with all the descendants of node $h'$
removed where $h'$ minimizes $g(h)$ among all nodes in tree $\hat{\psi}_m$, i.e., $h' = \argmin_{h \in I_{\hat{\psi}_m}} g(h)$. Set $m = m+1$.

\item[(c)] Repeat Step (b) until $\hat{\psi}_{m+1}$ consists only of the root node.
\end{enumerate}

\State Cross-validate to select the final tree from the candidate trees:

\begin{enumerate}
    \item[(a)] Fix the value of the penalization parameter $\lambda$ at some quantile of the asymptotic distribution of the splitting statistic defined in expression \eqref{eq: test_stat}. 
    
    \item[(b)] For each candidate tree $\hat{\psi}_m$, $\hat{\psi}_m \in \{\hat{\psi}_0, \hat{\psi}_1,\dots, \hat{\psi}_{\hat{M}}\}$, calculate the split complexity $G^{(\lambda)}(\hat{\psi}_m)$ given by equation \eqref{eq: split_complex} using the validation set by sending observations down the tree to calculate the splitting statistics defined in \eqref{eq: test_stat} for each internal node.
    
    \item[(c)] Select the final tree as the one that maximizes the validation set split complexity.
\end{enumerate}

\end{algorithmic}
\caption{\label{alg: git} Generalized Interaction Tree algorithm}
\end{algorithm}

\section{Subgroup-Specific Treatment Effect Estimators with Observational Data} \label{sec: estimators}

Implementation of the Generalized Interaction Tree algorithm requires specifying an estimator $\hat{\mu}_a(w)$ for $\mu_a(w) = \E[Y^{a}|\bm{X} \in w]$. In this section we define and discuss properties of three estimators for $\mu_a(w)$ that can be used with observational data. All theoretical results presented in this section assume that the partitioning is fixed (e.g.,~not data-dependent). In Sections \ref{sec:IPW}-\ref{sec:DR}, we consider the general case of treatment effect estimation for an arbitrary subgroup and in Section \ref{sec:CTA}, we discuss the use of these estimators in connection with the Generalized Interaction Tree algorithm.

\subsection{Identifiability of subgroup-specific treatment effects}
\label{sec:IPW}
The following conditions are sufficient to identify $\mu_a(w)$ using the observed data.
\begin{enumerate}
    \item Consistency of potential outcomes: $Y = Y^{1}A + Y^{0}(1-A)$. That is, an individual exposed to treatment $A = a$ has the observed outcome $Y$ equal to his or her potential outcome $Y^{a}$.
    \item Mean exchangeability: $\E[Y^{a}|\bm{X} = \bm{x}, A=a, \bm{X} \in w] = \E[Y^{a}|\bm{X}= \bm{x}, \bm{X} \in w]$ for all covariate patterns $\bm{x} \in w$ that have a positive density.
    \item Positivity: Each covariate pattern $\bm{x} \in w$ that has a positive density satisfies $0 < \PP(A = 1 | \bm{X} = \bm{x},  \bm{X} \in w) < 1$. 
\end{enumerate}
The following theorem shows that the subgroup-specific potential outcome means $\E[Y^{a}|\bm{X} \in w]$ are identifiable using the observed data (a similar arguement can be found in \cite{robertson2020Assessing}); we provide a proof in Web Appendix \ref{app: Proof-id}.
\begin{thm} \label{thm: identifiability}
Under identifibility conditions 1-3, the subgroup-specific potential outcome mean $\E[Y^{a}|\bm{X} \in w]$ can be written as the observed data functional
\begin{equation}
\label{gform-id}
\mu_a(w) = \E[\E[Y|\bm{X}, A = a]|\bm{X} \in w]
\end{equation}
or equivalently using the inverse probability weighting representation
\begin{equation}
\label{ipw-id}
\mu_a(w) = \frac{1}{\PP(\bm{X} \in w)} \E\left[ \frac{I(\bm{X} \in w, A = a)}{\PP(A = a|\bm{X})} Y\right].
\end{equation}
\end{thm}

\subsection{Inverse probability of treatment assignment based estimator} \label{sec: ipw}

Using plug-in estimators into identifiability result \eqref{ipw-id} gives the inverse probability weighting estimator, 
\begin{align} \label{eq: ipw1}
    \hat{\mu}_{\text{IPW}, a}(w) =\frac{1}{n(w)} \sum_{i = 1}^{n} \frac{I(\bm{X}_i \in w, A_i = a) Y_i}{e_a (\bm{X}_i; \hat{\bm{\beta}})}.
\end{align} 
Here, $\frac{n}{n(w)}$ is a non-parametric estimator for $\PP(\bm{X} \in w)^{-1}$, the inverse probability of being in subgroup $w$; and $e_a (\bm{X}; \hat{\bm{\beta}})$ is an estimator for $\PP(A = a|\bm{X})$, the probability of receiving treatment $a$ given covariates $\bm{X}$, indexed by the parameter $\bm{\beta}$. Throughout, we make the positivity assumption $e_a (\bm{X}; \hat{\bm{\beta}}) \geq \varepsilon >0$, for each $a \in \{0,1\}$.

The inverse probability weighting estimator $\hat{\mu}_{\text{IPW}, a}(w)$ is a weighted average of the outcome for all subjects in subgroup $w$ receiving treatment $a$, with weights being equal to the inverse of the estimated propensity scores. Consistency of $ \hat{\mu}_{\text{IPW}, a}(w)$ to the potential outcome mean follows from the identifiability expression \eqref{ipw-id} if $e_a (\bm{X}; \hat{\bm{\beta}})$ is correctly specified, that is $e_a (\bm{X}; \hat{\bm{\beta}})$ converges in probability to $\PP(A = a | \bm{X})$. 

Define the the true treatment effect difference between two disjoint subgroups denoted by $l$ and $r$ as
\begin{equation}
\label{T-def}
T(l,r) = (\mu_{1}(l) - \mu_{0}(l)) - (\mu_{1}(r) - \mu_{0}(r))
\end{equation}
The parameter $T(l,r)$ is the population value of the numerator of the splitting statistic \eqref{eq: test_stat}. Define $\hat T_{\text{IPW}}(l,r)$ by replacing the potential outcome means ($\mu_{a}(w), a \in \{0,1\}, w \in \{l,r\}$) in equation \eqref{T-def} by the corresponding inverse probability weighting estimators. 

The inverse probability weighting splitting statistic depends on an estimator of the variance of $\hat T_{\text{IPW}}(l,r)$. The following theorem provides the asymptotic variance of $\hat T_{\text{IPW}}(l,r)$ when the propensity score model is estimated using logistic regression, the most common model for estimating propensity scores in applied work. Although not explicit in the notation, interactions and higher-order terms may also be included in the model. 
\begin{thm}
\label{thm: ipw-var}
Assume that the propensity score is estimated using a correctly specified logistic regression model fit using the data in the union of the two disjoint subgroups $\{l,r\}$. Denote the estimated logistic regression coefficients as $\hat{\bm{\beta}}_{LR}$ and the corresponding asymptotic limit as $\bm{\beta}_{LR}$.
For a split into two subgroups $l$ and $r$ and any $w \in \{l,r\}$, define
$\bm{H}_{\bm{\beta}_{\text{LR}}, w} = \E\left[ \left(\frac{A Y}{e_1(\bm{X}; \bm{\beta}_{\text{LR}})} + \frac{(1-A) Y}{e_0(\bm{X}; \bm{\beta}_{\text{LR}})}\right) \frac{\partial}{\partial \bm{\beta}} e_1(\bm{X}; \bm{\beta})\bigg |_{\bm{\beta} = \bm{\beta}_{\text{LR}}}\bigg|\bm{X} \in w\right]$ and\\ $\mathbf{E}_{\bm{\beta}\bm{\beta}} = \E \left[ \left. \frac{\frac{ \partial}{\partial \bm{\beta}} e_1(\bm{X}; \bm{\beta})\bigg |_{\bm{\beta} = \bm{\beta}_{LR}} \left(\frac{\partial}{\partial \bm{\beta}} e_1(\bm{X}; \bm{\beta})\bigg |_{\bm{\beta} = \bm{\beta}_{LR}} \right)^T}{ e_1(\bm{X}; \bm{\beta}_{\text{LR}}) e_0(\bm{X}; \bm{\beta}_{\text{LR}}) } \right| \bm{X} \in \{l,r\}\right]$. The asymptotic variance of $\hat T_{IPW}(l,r)$ is given by
\begin{align}
    &\frac{1}{\PP(\bm{X} \in l|\bm{X} \in \{l,r\})} \left. \E\left( \frac{\left[Y A\right]^2 }{e_1(\bm{X}; \bm{\beta}_{\text{LR}})} + \frac{\left[Y (1-A)\right]^2 }{e_0(\bm{X}; \bm{\beta}_{\text{LR}})} \right| \bm{X} \in l\right) \nonumber \\
    &+ \frac{1}{\PP(\bm{X} \in r|\bm{X} \in \{l,r\})} \left. \E\left( \frac{\left[Y A \right]^2 }{e_1(\bm{X}; \bm{\beta}_{\text{LR}})} + \frac{\left[Y (1 - A)\right]^2 }{ e_0(\bm{X}; \bm{\beta}_{\text{LR}}) } \right| \bm{X} \in r\right) - T(l,r)^2 \nonumber  \\ 
    &- (\bm{H}_{\bm{\beta}_{\text{LR}}, l} - \bm{H}_{\bm{\beta}_{\text{LR}}, r})^T \mathbf{E}_{\bm{\beta}\bm{\beta}}^{-1} (\bm{H}_{\bm{\beta}_{\text{LR}}, l} - \bm{H}_{\bm{\beta}_{\text{LR}}, r}) \nonumber \\ 
    &- \frac{\left(\PP(\bm{X} \in r|\bm{X} \in \{l,r\})\left[\mu_{1}(l) - \mu_{0}(l)\right] + \PP(\bm{X} \in l|\bm{X} \in \{l,r\}) \left[\mu_{1}(r) - \mu_{0}(r)\right] \right)^2}{\PP(\bm{X} \in l|\bm{X} \in \{l,r\}) \PP(\bm{X} \in r|\bm{X} \in \{l,r\})}. \label{eq:var-ipw}
\end{align}
\end{thm}
We present a proof and give a consistent variance estimator in Web Appendix \ref{app: ipw_var}.

The asymptotic variance result in Theorem \ref{thm: ipw-var} is an extension of the results on marginal estimators from \cite{lunceford2004stratification}, to subgroup-specific treatment effect estimators. The variance \eqref{eq:var-ipw} consists of five terms. The first three terms represent the variance if both the propensity score model and $\PP(\bm{X} \in l|\bm{X} \in \{l, r\})$ were known. The fourth term reflects the variance reduction when using the estimated propensity score compared with using the true propensity score (a similar observation was made in \cite{lunceford2004stratification} for the non-subgroup case). Interestingly, the last term in \eqref{eq:var-ipw} shows that using a non-parametric estimator for $\PP(\bm{X} \in l|\bm{X} \in \{l, r\})$ results in further variance reduction compared with using the true (unknown) conditional probability.

\subsection{G-formula estimator}

Using plug-in estimators into identifiability result \eqref{gform-id} gives the g-formula estimator
\begin{align*}
    \hat{\mu}_{\text{G}, a}(w) = \frac{1}{n(w)} \sum_{i=1}^n I(\bm{X}_i \in w) g_a(\bm{X}_i; \hat{\bm{\eta}}_a),
\end{align*}
where $g_a(\bm{X}; \hat{\bm{\eta}}_a)$ is an estimator for $\E[Y|\bm{X}, A=a]$ and $\bm{\eta}_a$ denotes the parameter used to estimate the model. It follows from identifibility result \eqref{gform-id} that if the outcome model is correctly specified (i.e.,~$g_a(\bm{X}; \hat{\bm{\eta}}_a)$ converges in probability to $\E[Y|\bm{X}, A = a]$), $\hat{\mu}_{\text{G}, a}(w)$ is a consistent estimator for $\mu_a(w)$. A variance estimator for $\hat{\mu}_{\text{G}, a}(w)$ is given by equation (13) in \cite{steingrimsson2019subgroup}.

\subsection{Doubly robust estimator}
\label{sec:DR}

Web Appendix \ref{app: IF_DR} shows that the first order influence function \citep{van2003unified} of $\mu_a(w)$ under the non-parametric model is given by
\begin{align*}
     \psi_a(w) = \frac{1}{\PP(\bm{X}\in w)}\bigg\{ & I(\bm{X}\in w) (\E(Y|\bm{X}, A = a) - \mu_a(w)) \\ 
     & + \frac{I(\bm{X}\in w, A = a)}{\PP(A=a|\bm{X})} \left[Y - \E(Y|\bm{X}, A = a)\right] \bigg\} .
\end{align*}
More precisely, we show that $\psi_a(w)$ is a mean zero function that solves the equation $\frac{\partial \psi_{a,t}(w)}{\partial t}\bigg|_{t=0} =  \E[\psi_a(w) l(O)]$, where $l(O)$ is the score of the observed data $O = (\bm{X},A,Y)$ and the left hand side of the equation is the pathwise derivative of the target parameter w.r.t.~a set of parametric submodels indexed by $t$.

This influence function suggests the estimator 
\begin{align}\label{eq: dr_est}
    \hat{\mu}_{\text{DR, a}}(w) =& \frac{1}{n(w)} \sum_{i = 1}^n \left(I(\bm{X}_i \in w)  g_a(\bm{X}_i; \hat{\bm{\eta}}_a) +  \frac{I(\bm{X}_i \in w, A_i = a) }{ e_a(\bm{X}_i; \hat{\bm{\beta}}) } \left[Y_i - g_a(\bm{X}_i; \hat{\bm{\eta}}_a) \right] \right).
\end{align}
Following \cite{van1996weak}, for a function $f(O)$ define $\mathbb{P}_n(f(O)) = \frac{1}{n} \sum_{i=1}^n f(O_i)$ and $\mathbb{G}_n(f(O)) = \sqrt{n}(\mathbb{P}_n[f(O)] - \E[f(O)])$. For any functions $e_a (\bm{X}; \bm{\beta})$, $g_a(\bm{X}; \bm{\eta}_a)$, and $\gamma$, define
\begin{align*}
    H(e_a(\bm{X}; \bm{\beta}), g_a(\bm{X}; \bm{\eta}_a), \gamma) =& \gamma I(\bm{X} \in w) \left(\vphantom{\frac{1}{n}} g_a(\bm{X}; \bm{\eta}_a) + \frac{I(A = a)}{e_a(\bm{X}; \bm{\beta})} (Y - g_a(\bm{X}; \bm{\eta}_a) )\right).
\end{align*} 
Using this notation
\[
\mathbb{P}_n\left(H\left( e_a(\bm{X}; \hat{\bm{\beta}}), g_a(\bm{X}; \hat{\bm{\eta}}_a), \hat \gamma = \frac{n}{n(w)}\right)\right) = \hat{\mu}_{\text{DR,a}}(w).
\]

Let $\bm{\beta}^*$ and $\bm{\eta}^*_a$ be the asymptotic limits of $\hat{\bm{\beta}}$ and $\hat{\bm{\eta}}_a$ (assumed to exist). By the assumptions already made, $\hat \gamma$ is a consistent estimator for $\PP(\bm{X} \in w)^{-1}$. To derive the large sample properties of $\hat{\mu}_{\text{DR,a}}(w)$, we make the following assumptions:
\begin{itemize}
\item[A.1] The process $H(e_a(\bm{X}; \hat{\bm{\beta}}), g_a(\bm{X}; \hat{\bm{\eta}}_a), \hat \gamma)$ and the limit $H(e_a(\bm{X}; \bm{\beta}^*), g_a(\bm{X}; \bm{\eta}^*_a), \PP(\bm{X} \in w)^{-1})$ are Donsker \citep{van1996weak}.
\item[A.2] $||H( e_a(\bm{X}; \hat{\bm{\beta}}), g_a(\bm{X}; \hat{\bm{\eta}}_a), \hat \gamma) - H( e_a(\bm{X}; \bm{\beta}^*), g_a(\bm{X}; \bm{\eta}^*_a), \PP(\bm{X} \in w)^{-1})||_2 \xrightarrow{p} 0$,
where "$\xrightarrow{p}$" denotes convergence in probability.
\item[A.3] $\E[H(e_a(\bm{X}; \bm{\beta}^*), g_a(\bm{X}; \bm{\eta}^*_a), \PP(\bm{X} \in w)^{-1})^2] < \infty$.
\item[A.4] At least one of the following holds:
\begin{align*}
g_a(\bm{X}; \hat{\bm{\eta}}_a) \xrightarrow{p} \E[Y|\bm{X}, A=a] \mbox{\quad or \quad} e_a(\bm{X}; \hat{\bm{\beta}}) \xrightarrow{p} \PP(A=a |\bm{X}).
\end{align*}
\end{itemize}
\begin{thm}
\label{Thm-DR}
Under Assumptions A.1-A.4, we have that:
\begin{enumerate}
\item The doubly robust estimator is consistent, that is $\hat{\mu}_{\text{DR,a}}(w) \xrightarrow{p} \mu_a(w)$.
\item The doubly robust estimator has rate of convergence
\begin{align}
||\hat{\mu}_{\text{DR,a}}(w) - &\mu_a(w)||_2  = \label{rate} \\ &O_{P}\left(\frac{1}{\sqrt{n}} + \big|\big| e_a(\bm{X}; \hat{\bm{\beta}}) - \PP(A=a|\bm{X})\big|\big|_2 \big|\big| g_a(\bm{X}; \hat{\bm{\eta}}_a) - \E[Y|\bm{X}, A=a]\big|\big|_2 \right) \nonumber
\end{align}
\end{enumerate}
\end{thm}
We provide a proof in Web Appendix \ref{app: DR_Inference}. Theorem \ref{Thm-DR} gives useful insights into the asymptotic behaviour of the doubly robust estimator. Assumption A.4 implies that the estimator is doubly robust in that it is consistent if at least one of the models
$g_a(\bm{X}; \hat{\bm{\eta}}_a)$ or $e_a(\bm{X}; \hat{\bm{\beta}})$ are consistent, but not necessarily both. The rate of convergence result given by equation \eqref{rate} shows that if the combined rate of convergence of 
$g_a(\bm{X}; \hat{\bm{\eta}}_a)$ and $e_a(\bm{X}; \hat{\bm{\beta}})$ to the true parameters $\E[Y|\bm{X}, A=a]$ and $\PP(A=a |\bm{X})$ is at least $\sqrt{n}$, then $\hat{\mu}_{\text{DR,a}}(w)$ is $\sqrt{n}$ consistent.
The Donsker requirements in assumption A.1 restrict the entropy of the estimators 
$e_a(\bm{X}; \hat{\bm{\beta}})$ and $g_a(\bm{X}; \hat{\bm{\eta}}_a)$. Although substantially more flexible than requiring a parametric $
\sqrt{n}$ rate of convergence, many modern machine learning methods do not satisfy this assumption. To overcome that, cross-fitting can be used \citep{chernozhukov2018double,robins2008higher}. 

\subsection{Causal Interaction Tree Algorithms}
\label{sec:CTA}
We define the Inverse Probability Weighting, G-formula, and Doubly Robust Causal Interaction Tree algorithms (IPW-CIT, G-CIT, and DR-CIT, respectively) as the Generalized Interaction Tree algorithm implemented using $\hat \mu_{IPW,a}(w)$, $\hat \mu_{G,a}(w)$, $\hat \mu_{DR,a}(w)$. Collectively, we refer to these three algorithms as the Causal Interaction Tree (CIT) algorithms.

All the theory developed in this section is for a fixed partitioning, but the Causal Interaction Tree algorithm is based on data-dependent partitionings. Dealing with fixed partitions is a standard simplification made when dealing with theoretical properties of estimators derived from the Classification and Regression Tree algorithm \citep[Ch. 9.3]{breiman1984classification}, and to the best of our knowledge all theory for single trees relies on simplifying assumptions such as assuming all covariates are binary or only focusing on the first step of the tree building process (i.e.,~ignoring the pruning step). Nevertheless, because all decision-making including splitting decision, pruning, and final tree selection in the Causal Interaction Tree algorithms is driven by the estimators for $\mu_a(w)$, we expect that more robust and efficient estimators for $\mu_a(w)$ will lead to better performance of the algorithms. For censored observations, more efficient and robust estimators for the statistics used for decision-making in tree-based algorithms have been shown to improve empirical performance compared to naive estimators \citep{steingrimsson2016doubly, steingrimsson2019censoring}. In Section \ref{sec: simulations}, we will explore the performance of the Causal Interaction Tree algorithms in simulations when using different levels of misspecification of the propensity score and/or outcome model.

\section{Simulations} \label{sec: simulations}

\subsection{Simulation setup} \label{sec: sim_setup}

The covariate vector was simulated from a 6-dimensional mean zero multivariate normal distribution, where $\text{cov}(X^{(j)}, X^{(k)}) = 0.3$ when $j \ne k$, and $\text{Var}(X^{(j)}) = 1$. The treatment indicator $A$ was simulated from a $\text{Bernoulli}(p)$ distribution, where $p = \text{expit} \left( 0.6X^{(1)} -0.6X^{(2)} + 0.6X^{(3)}  \right)$. The outcome was simulated using two different settings:
\begin{itemize}
    \item $Y = 2 + 2A + 2I\left(X^{(1)} < 0\right) + \exp\left(X^{(2)}\right) + 3 I\left(X^{(4)} > 0\right) + \left(X^{(5)}\right)^3 + \epsilon$, where $\epsilon \sim N(0, 1)$. For this setting, the treatment effect is the same for all covariate values and the correct tree consists only of the root node. We refer to this simulation setting as the homogeneous treatment effect setting.

    \item $Y = 2 + 2A + 2I\left(X^{(1)} < 0\right) + \exp\left(X^{(2)}\right) + 3A I\left(X^{(4)} > 0\right) + \left(X^{(5)}\right)^3 + \epsilon$, where $\epsilon \sim \text{N}(0, 1)$. For this setting, the treatment effect differs depending on whether $X^{(4)} >0$ or not and the correct tree splits on $X^{(4)}$ at 0. We refer to this simulation setting as the heterogeneous treatment effect setting.
\end{itemize}
For both simulation settings, a training and a test set were generated by drawing 1000 independent samples from the joint distribution of $(\bm{X}, A, Y)$.

\subsection{Evaluation measures}

We used the following measures to evaluate the performance of different methods:
\begin{itemize}
    \item Mean Squared Error (MSE): Let $\hat{\rho}(\bm{X}_i)$ be a prediction for $\rho(\bm{X}_i) = \E[Y|A=1,\bm{X}_i] - \E[Y|A=0,\bm{X}_i]$. The mean squared error is defined as $1000^{-1} \sum_{i=1}^{1000} (\hat \rho(\bm{X}_i) - \rho(\bm{X}_i))^2$, where $\bm{X}_i, i = 1, \ldots, 1000$ are the covariates from the test set.
    
    \item Proportion of Correct Trees: Splitting on a continuous covariate at the correct split point has a probability of zero. Therefore, a tree is defined to be correct if it splits on all the continuous variables the correct number of times independently of the selection of the splitting point and if it splits on all the categorical or ordinal variables at the correct split points.
    
    \item Number of Noise Variables: The average number of times the tree splits on one of the noise variables (i.e., $\{X^{(1)}, \dots, X^{(6)}\}$ for the homogeneous treatment effect setting and $\{X^{(1)}, X^{(2)}, X^{(3)}, X^{(5)}, X^{(6)}\}$ for the heterogeneous treatment effect setting).
    
    \item Pairwise Prediction Similarity: Let $I_T(i,j)$ and $I_M(i,j)$ be indicators if participants $i$ and $j$ fall in the same terminal node when running down the true tree and the fitted tree, respectively. Pairwise prediction similarity is defined as
\[
1 - \sum_{i=1}^{1000} \sum_{j>i}^{1000} \frac{|I_T(i,j) - I_M(i,j)|}{\binom{1000}{2}}
\] 
and it measures the ability of the tree-based algorithms to stratify observations into different groups.

    \item Proportion of Correct First Splits: The proportion of fully grown trees $\psi_{\max}$ that make a correct first split (only applicable to the heterogeneous simulation setting). 
\end{itemize}


\subsection{Implementation}
\label{sec: implementation}
We implemented the large tree $\psi_{\max}$ for the Causal Interaction Tree algorithms, IPW-CIT, G-CIT, and DR-CIT, using \texttt{rpart}'s ability to accommodate user written splitting and evaluation functions. 

The implementation required choosing what part of the data is used to fit the propensity score and outcome models in the estimators (e.g.,~fit a single model in the parent node or fit separate models in the child nodes). The results presented in Section \ref{sec: sim-results} use models that were fit using the data in the parent node being considered for splitting. In Web Appendix \ref{app: sim_out_insplt} we present simulation results when models were fit using the whole dataset prior to the tree building process and separately using the data in each of the potential child nodes.

To evaluate the impact of misspecifying the propensity score and outcome models, we implemented the tree-based algorithms using the correct model specification, a version that uses a misspecified functional form of the covariates and a version that has unmeasured common cause of the outcome and treatment assignment. The correct logistic regression model for estimating the propensity scores in $\hat{\mu}_{\text{IPW}, a}(w)$ and $\hat{\mu}_{\text{DR}, a}(w)$ includes main effects of $X^{(1)}, X^{(2)}$ and $X^{(3)}$. The correct linear regression outcome model used to implement $\hat{\mu}_{\text{G}, a}(w)$ and $\hat{\mu}_{\text{DR}, a}(w)$ includes $A$, $I\left(X^{(1)}<0\right)$, $\exp\left(X^{(2)}\right)$, $I\left(X^{(4)}>0\right)$ and $\left(X^{(5)}\right)^3$ for the homogeneous treatment effect setting and an interaction between $A$ and $I\left(X^{(4)}>0\right)$ instead of $I\left(X^{(4)}>0\right)$ for the heterogeneous setting.

The first form of model misspecification corresponds to including incorrect functional forms of covariates.
For the outcome model, main effects of treatment and covariates and all two-way treatment-covariate interactions are included in their original form. For the propensity score model, exponentiated forms of all covariates are included.

The second form of model misspecification mimics the scenario where there is unmeasured common cause of the outcome and treatment assignment. For that setting, we exclude $X^{(2)}$ from that data. The outcome model includes main effect of treatment and all covariates except for $X^{(2)}$ and all two-way treatment-covariate interactions that do not involve $X^{(2)}$. The propensity score model includes main effects of all covariates except for $X^{(2)}$. When there are unmeasured common causes of outcome and treatment the mean exchangeabilty condition is expected to be violated and all estimators are expected to be biased.

For the final tree selection step, the training dataset was split into an initial tree building dataset of size 800 and a validation set of the remaining 200 observations. The penalization parameter $\lambda$ was chosen to be the 95th percentile of a $\chi^2$-distribution with 1 degree of freedom, which is the limiting distribution of the splitting statistic defined in \eqref{eq: test_stat} when the treatment effect is identical in the two subgroups.

We compared the performance of the Causal Interaction Tree algorithms with the Causal Tree (CT) algorithm proposed by \cite{athey2016recursive}. To apply the CT algorithm to datasets where the treatment is not randomly assigned, we set the \texttt{weights} parameter to the inverse of the observation-specific propensity scores estimated from the whole dataset. The propensity score model was implemented in the same way as for the CIT algorithms. 

Implementation of the Causal Tree algorithm using the \texttt{R} package \texttt{causalTree} required selecting several tuning parameters. Following \cite{athey2016recursive}, we set both the splitting rule (\texttt{split.Rule}) and the cross-validation method (\texttt{"cv.option"}) to \texttt{"CT"}. We also set \texttt{split.Honest} and \texttt{cv.Honest} to \texttt{TRUE} for honest splitting and cross-validation. We refer to this setting as "Original CT". In addition, we also compared the performance of the Causal Interaction Tree algorithms against a Causal Tree algorithm with the combination of tuning parameters that has the highest rank on average in terms of minimizing MSE across the two simulation settings. Simulations presented in Web Appendix \ref{app: ct_sim} show that the parameter setting that has the highest average rank is setting \texttt{split.Rule} to \texttt{"tstats"} with the honest version (\texttt{split.Honest = TRUE}) and \texttt{cv.option} to \texttt{"matching"}.  We refer to this parameter combination as "Best CT" in simulation results. We refer to Web Appendices \ref{app: ct_sim} and \ref{app: ct_nohonest} for further details on implementation of the Causal Tree algorithms. Code implementing simulations presented in this section is available from \url{github.com/jiabei-yang/CIT}.

\subsection{Simulation results}
\label{sec: sim-results}

\begin{figure}[htbp]
        \begin{center}
        \centerline{\includegraphics[width = 6in]{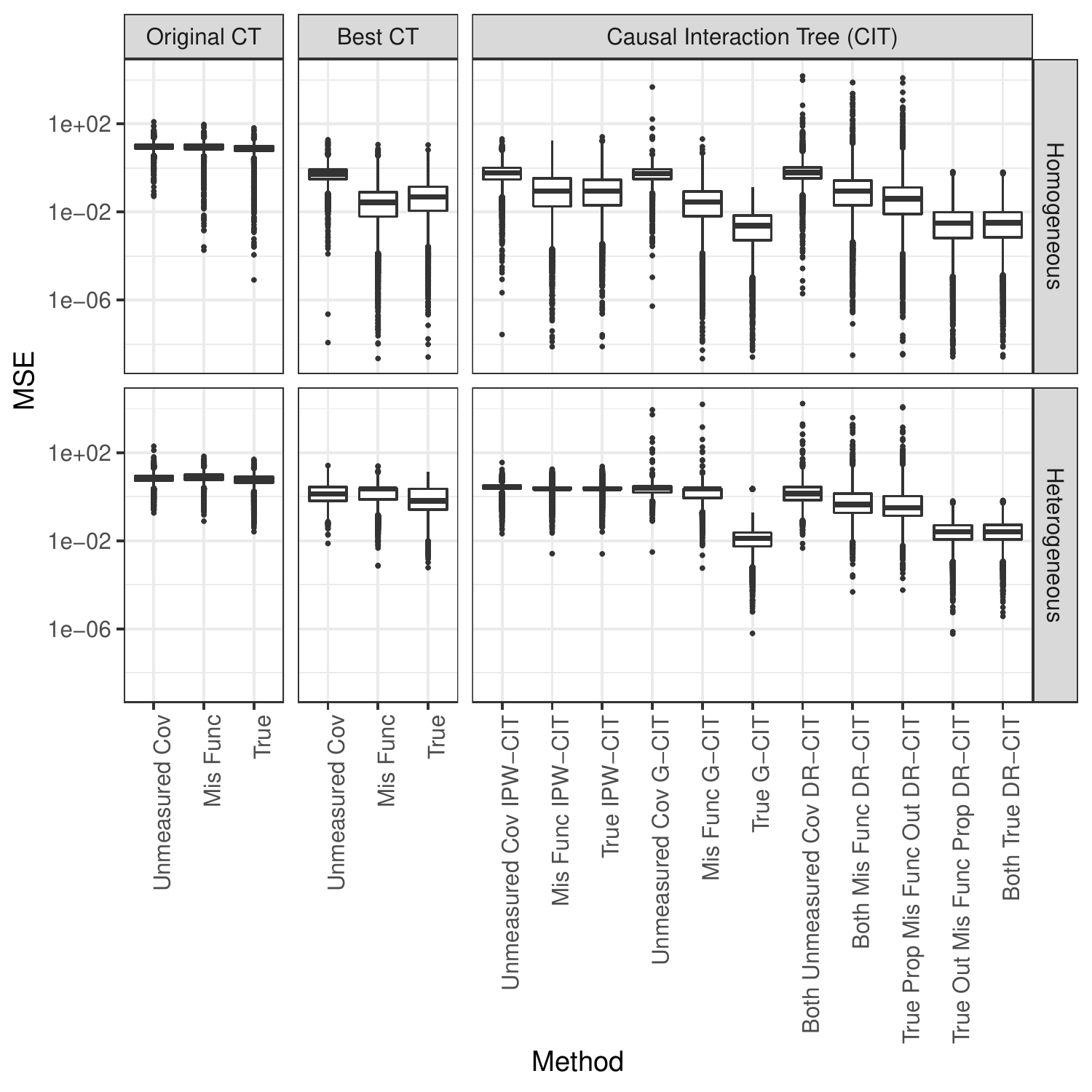}}
    \end{center}
    \caption{Mean squared error (MSE) for the different tree building algorithms for the homogeneous (top) and the heterogeneous (bottom) settings described in Section \protect\ref{sec: sim_setup}. Lower values indicate better performance. "Original CT" refers to the original Causal Tree algorithm by \protect\cite{athey2016recursive}. "Best CT" refers to the Causal Tree algorithm with optimized splitting rule and cross-validation method. IPW-CIT, G-CIT and DR-CIT refer to the Inverse Probability Weighting, G-formula, and Doubly Robust Causal Interaction Tree algorithms, respectively. 
    "Unmeasured Cov" refers to having an unmeasured common cause of outcome and treatment assignment as described in Section \protect\ref{sec: sim_setup}. "Mis Func" refers to using a misspecified functional form of the covariates in the propensity score and/or outcome model as described in Section \protect\ref{sec: sim_setup}. "True" refers to using correctly specified propensity score and/or outcome models. For DR estimators, "Prop" and Out" stand for the propensity score model and outcome model, respectively.}
    \label{fig: mse}
\end{figure}

We used 10,000 simulations for both settings described in Section \ref{sec: sim_setup} to compare the performance of Causal Interaction Trees to the Causal Tree algorithms, implemented as described in Section \ref{sec: implementation}. Figure \ref{fig: mse} shows boxplots of MSE and Table \ref{tab: summ_stat_cv1} shows the proportion of correct trees, average number of noise variables, and pairwise prediction similarity for both simulation settings, and the proportion of trees making a correct first split in the heterogeneous setting.  

\begin{table}[htbp]
\centering 
\footnotesize
\begin{tabular}{@{}llccccccccccccc@{}}
\toprule
& &\multicolumn{3}{c}{Homogeneous Effect} &  & \multicolumn{4}{c}{Heterogeneous Effect}\\
\cmidrule[0.5pt]{3-5} \cmidrule[0.5pt]{7-10}
 & & Correct & Number& &  & Correct & Number & & Correct\\
Algorithm & Model & Trees & Noise & PPS & & Trees & Noise & PPS & First Split \\ 
  \hline
   Original CT &  Unmeasured Cov & 0.00 & 28.36 & 0.05 &  & 0.01 & 21.41 & 0.55 & 0.63 \\ 
   &  Mis Func & 0.00 & 25.28 & 0.07 &  & 0.00 & 19.85 & 0.55 & 0.30 \\ 
   &  True & 0.02 & 25.91 & 0.08 &  & 0.02 & 19.31 & 0.57 & 0.58 \\ 
   Best CT &  Unmeasured Cov & 0.99 & 0.02 & 1.00 &  & 0.52 & 0.15 & 0.77 & 0.87 \\ 
   &  Mis Func & 1.00 & 0.01 & 1.00 &  & 0.28 & 0.29 & 0.66 & 0.47 \\ 
   &  True & 1.00 & 0.01 & 1.00 &  & 0.58 & 0.19 & 0.81 & 0.84 \\ 
  IPW-CIT &  Unmeasured Cov & 0.90 & 0.50 & 0.95 &  & 0.11 & 0.50 & 0.57 & 0.71 \\ 
   & Mis Func & 0.78 & 0.82 & 0.90 &  & 0.03 & 0.93 & 0.53 & 0.20 \\ 
   & True & 0.90 & 0.57 & 0.95 &  & 0.06 & 0.68 & 0.53 & 0.47 \\ 
  G-CIT &  Unmeasured Cov & 0.95 & 0.11 & 0.97 &  & 0.27 & 0.12 & 0.61 & 0.96 \\ 
   & Mis Func & 0.95 & 0.10 & 0.97 &  & 0.30 & 0.10 & 0.63 & 0.96 \\ 
   & True & 1.00 & 0.00 & 1.00 &  & 0.99 & 0.00 & 1.00 & 1.00 \\ 
  DR-CIT &  Both Unmeasured Cov & 0.88 & 0.77 & 0.93 &  & 0.53 & 0.70 & 0.78 & 0.97 \\
   & Both Mis Func & 0.89 & 0.75 & 0.94 &  & 0.65 & 0.86 & 0.85 & 0.99 \\ 
   & True Prop Mis Func Out & 0.89 & 0.78 & 0.93 &  & 0.70 & 0.73 & 0.87 & 1.00 \\ 
   & True Out Mis Func Prop & 0.95 & 0.14 & 0.98 &  & 0.93 & 0.12 & 0.99 & 1.00 \\ 
   & Both True & 0.95 & 0.13 & 0.98 &  & 0.94 & 0.13 & 0.99 & 1.00 \\ 
\bottomrule
\end{tabular} 
\caption{Proportion of correct trees (higher is better), average number of noise variables used for splitting (lower is better), pairwise prediction similarity (higher is better), 
and proportion of trees making the first split correctly (only for heterogeneous setting, higher is better). Columns 3, 4 and 5 show simulation results when the treatment effect is homogeneous, and columns 6, 7, 8 and 9 show simulation results when the treatment effect is heterogeneous. "Original CT" refers to the original Causal Tree algorithm by \protect\cite{athey2016recursive}. "Best CT" refers to the Causal Tree algorithm with optimized splitting rule and cross-validation method. IPW-CIT, G-CIT and DR-CIT refer to the Inverse Probability Weighting, G-formula, and Doubly Robust Causal Interaction Tree algorithms, respectively. "Unmeasured Cov" refers to having an unmeasured common cause of outcome and treatment assignment as described in Section \protect\ref{sec: sim_setup}. "Mis Func" refers to using a misspecified functional form of the covariates in the propensity score and/or outcome model as described in Section \protect\ref{sec: sim_setup}. "True" refers to using correctly specified propensity score and/or outcome models. For DR estimators, "Prop" and Out" stand for the propensity score model and outcome model, respectively. \label{tab: summ_stat_cv1}}
\end{table}

The results in Figure \ref{fig: mse} and Table \ref{tab: summ_stat_cv1} are consistent with what is expected based on the properties of the subgroup-specific treatment effect estimators described in Section \ref{sec: estimators}. When the propensity score and/or outcome models are correctly specified, the G-CITs show the best overall performance, closely followed by the DR-CITs, and the IPW-CITs perform worse than their peers.
All Causal Interaction Tree algorithms show the best performance when the correct outcome model and/or propensity score model required for implementation are used. When the models are misspecified, the robustness of the doubly robust estimator is also confirmed: a) the DR-CITs with a correctly specified outcome model but a misspecified propensity score model perform substantially better than the IPW-CITs using a misspecified propensity score model, and b) the DR-CITs with a correctly specified propensity score model but a misspecified outcome model perform similarly to the G-CITs with a misspecified outcome model in the homogeneous setting but substantially better in the heterogeneous simulation setting. Overall, the DR-CITs have the best performance, followed by the G-CITs; the IPW-CITs show the worst overall performance among the Causal Interaction Tree algorithms. Finally, the results in Figure \ref{fig: mse} and Table \ref{tab: summ_stat_cv1} show that G-CITs and DR-CITs perform substantially better than both versions of the Causal Tree algorithms.
 
To further evaluate the performance of the Causal Interaction Tree algorithms, Web Appendix \ref{app: add_sim} includes the following additional simulation results:
\begin{itemize}

    \item Web Appendix \ref{app: cv1_time} compares the running time of the Causal Tree and Causal Interaction Tree algorithms. Somewhat surprisingly, 
    the DR-CITs run on average 8-30 times faster than the IPW-CITs and the G-CITs. The reason is that unlike the variance estimator for the other two Causal Interaction Tree algorithms, the variance estimator of the doubly robust estimator does not involve calculating the inverse of the quadratic form of the design matrix, substantially reducing the computational complexity.
    
    \item Web Appendix \ref{app: sim_fts2} presents results when an alternative method for final tree selection proposed in \cite{steingrimsson2019subgroup} is used in connection with the Causal Interaction Tree algorithms. In short, the method uses the prediction from a random forest algorithm as a surrogate for the truth when performing cross-validation. Hence, it relies on the assumption that the random forest algorithm predictions are more accurate than the single tree based predictions. A description of this final tree selection method is given in Web Appendix \ref{app: FTS2}. The results show that the Causal Interaction Tree algorithms are more likely to overfit when the alternative final tree selection method is used.

    \item The simulations presented in this section use propensity score and outcome models that are fit using data in the parent node. Alternatives include fitting the models prior to the tree building process or fitting separate models in each of the child nodes for each possible split. In Web Appendix \ref{app: ipw_var_insplt} we present the analogous result to Theorem \ref{thm: ipw-var} when separate models are fit in each of the child nodes. 
    
    Web Appendix \ref{app: sim_out_insplt} presents simulation results for Causal Interaction Tree algorithms when the propensity score and the outcome models are fitted a) prior to the tree building process using the whole dataset and b) within each child node separately for each possible split. The results when the models are fitted using the whole dataset are similar to those when the models are fitted using the data in the node that is being considered for splitting. When the models are fitted separately in each child node, the Causal Interaction Tree algorithms perform in general worse than when the models are fitted using the data in the parent node.
     

    \item Web Appendix \ref{app: sim_bin_mixed} presents simulations when the outcome is binary and the covariate vector includes both continuous and categorical variables. The results show similar trends to the simulations presented in this section.
 \end{itemize}

 


\section{Analysis of the SUPPORT study}
\label{sec:data-analysis}

We used the Causal Interaction Tree algorithms to analyze data from the Study to Understand Prognoses and Preferences for Outcomes and Risks of Treatments (SUPPORT), an observational study that evaluated the effectiveness of right heart catheterization (RHC) on critically ill patients \citep{connors1996effectiveness}. The original analysis of the SUPPORT data \citep{connors1996effectiveness} used matching to analyze the data; a follow-up analysis by \cite{hirano2001estimation} used inverse probability weighting to estimate the average treatment effect. At the time of submission, the data is publicly available at \url{http://biostat.mc.vanderbilt.edu/wiki/Main/DataSets}. 

The dataset contains information on 5735 participants; the treatment group (2184 participants) received RHC during the first 24 hours in the intensive care unit; the control group (3551 participants) did not receive RHC in the same time window. Participants who experienced death within 48 hours were excluded from the original study. We focus on 30-day survival as the outcome of interest; there was no censoring so we treat the outcome as binary. There are 51 covariates available for analysis; a list of the covariates is included in Web Appendix \ref{app: data_analysis}.
For further details on the study, we refer to \cite{connors1996effectiveness}. 

We applied the three Causal Interaction Tree algorithms described in Section \ref{sec: simulations} to the SUPPORT data; for comparison, we also implemented the ``orignal'' and ``best'' Causal Tree algorithms. Following \cite{connors1996effectiveness}, we modeled the propensity score using a logistic regression model that included the main effects of all the covariates. The outcome model was fit using the main effects of treatment and all covariates, and all two-way treatment-covariate interactions. We fit the propensity score and outcome models using the data in the node that was being considered for splitting. The initial tree building dataset was a random sample of size $4588$ ($80\%$ of the original dataset); the remaining dataset of size $1147$ ($20\%$ of the original dataset) was used as the validation set for final tree selection. Other implementation choices were the same as those in the simulations described in Section \ref{sec: simulations}. 

The final tree from all three Causal Interaction Tree algorithms and the "best" Causal Tree algorithm consists only of a root node. That is, none of the algorithms identified any subgroups with differential treatment effects. The root-only tree is consistent with the results in \cite{connors1996effectiveness}, where none of the pre-defined subgroups were found to be associated with a larger treatment effect.
On the contrary, the "original" Causal Tree algorithm produced a large tree with 232 terminal nodes. That the "original" Causal Tree algorithm builds a large tree and that the "best" Causal Tree algorithm identifies no subgroups are consistent with the simulations where in both the homogeneous and heterogeneous settings, the former tends to build larger trees than the true tree and the latter tends to identify no subgroups (see Table \ref{tab: summ_stat_cv1}).

The large tree $\psi_{\max}$ (from step 1 of the Generalized Interaction Tree algorithm prior to pruning and final tree selection) built by the Doubly Robust Causal Interaction Tree (DR-CIT) first splits on if the SUPPORT model estimate of the probability of surviving 2 months at study entry is greater than or equal to 0.85 or not. The treatment effect for those in the node with 2-month survival probability greater than 0.85 was estimated to be 0.26 with a 95\% bootstrap interval of $[-0.09, 0.39]$; for those with 2-month survival probability lower than 0.85, the treatment effect was estimated to be 0.08 with a 95\% bootstrap interval $[0.04, 0.10]$. The bootstrap intervals were calculated using the nonparametric bootstrap with 1000 bootstrap samples assuming a fixed split and should therefore be considered exploratory. 
Therefore, the point estimate of the treatment effect of RHC is larger among those with higher 2-month survival probability on study entry although the difference IS not statistically significant. \cite{connors1996effectiveness} performed a pre-defined subgroup analysis based on the probability of surviving 2 months at study entry. In agreement with our results, they found that patients with a predicted 2-month survival probability greater than 0.60 tended to have increased effect of RHC on death although the difference was not statistically significant.

\section{Discussion}

The Causal Interaction Trees (CIT) algorithms are novel methods for subgroup identification using observational data. They are extensions of the interaction tree algorithm that utilize inverse probability weighting, g-formula, or doubly robust subgroup-specific treatment effect estimators for decision-making during tree construction. The consistency of the first two estimators requires correct specification of the propensity score or the outcome model, respectively. The doubly robust estimator only requires one of the propensity score and the outcome model to be correctly specified, but not necessarily both. We evaluated the finite sample properties of the three algorithms in simulations and implemented them to analyze data from an observational study evaluating the effectiveness of right heart catheterization on critically ill patients.

Ensemble-based methods that average multiple trees, such as bagging or random forest, usually improve prediction accuracy over single trees. Single tree structures have the advantage of partitioning the covariate space into identifiable subsets with differential treatment effects, and therefore construct an interpretable treatment effect stratification rule. In contrast, ensemble methods that average multiple trees result in black box prediction models that do not provide interpretable treatment effect stratification. Future research should consider using Causal Interaction Trees as building blocks for ensemble methods. Further extensions to address more complex data structures, such as censored or longitudinal data, may also prove useful in practice \citep{wei2020precision}.

\newpage 

\FloatBarrier

\section*{{\bf  Supplementary Web Appendix}}

\setcounter{section}{0}
\renewcommand{\thesection}{S.\arabic{section}}
\renewcommand{\theHsection}{Supplement.\thesection}

\setcounter{equation}{0}
\renewcommand{\theequation}{S-\arabic{equation}}
\renewcommand{\theHequation}{Supplement.\theequation}

\setcounter{table}{0}
\renewcommand{\thetable}{S-\arabic{table}}
\renewcommand{\theHtable}{Supplement.\thetable}

\setcounter{figure}{0}
\renewcommand{\thefigure}{S-\arabic{figure}}
\renewcommand{\theHfigure}{Supplement.\thefigure}

\setcounter{page}{1}
\renewcommand{\thepage}{\arabic{page}}

References to figures, tables, theorems and equations preceded by ``S-'' are internal to this supplement; all other references refer to the main paper.

\section{Properties of the subgroup-specific treatment effect estimators} \label{app: ipw}

\subsection{Proof of Identifiability of subgroup-specific treatment effect estimators}
\label{app: Proof-id}

\textbf{Proof of Theorem \ref{thm: identifiability}:} We start by showing the identifiability result \eqref{gform-id}
\begin{align*}
\E[Y^{a}|\bm{X} \in w] &= \E\left[\E\left[Y^{a}|\bm{X}\right]\bigg| \bm{X} \in w\right] \\
&= \E\left[\E\left[Y^{a}|\bm{X}, A = a\right]\bigg| \bm{X} \in w \right] \\
&= \E\left[\E\left[Y|\bm{X}, A = a\right]\bigg|\bm{X} \in w \right] 
\end{align*}
For results \eqref{ipw-id}, we have
\begin{align*}
\E[Y^{a}|\bm{X} \in w] &= \E\left[\E\left[Y|\bm{X}, A = a\right]\bigg|\bm{X} \in w \right] \\
&= \E\left[\E\left[\frac{I( A = a)}{\PP(A =a|\bm{X})}Y |\bm{X}\right]\bigg|\bm{X} \in w \right] \\
&=\E\left[\frac{I(\bm{X} \in w)}{\PP(\bm{X} \in w)} \E\left[\frac{I( A = a)}{\PP(A =a|\bm{X})}Y |\bm{X}\right]\right] \\
&= \frac{1}{ \PP(\bm{X} \in w)} \E\left[\E\left[\frac{I(\bm{X} \in w, A = a)}{\PP(A =a|\bm{X})}Y |\bm{X}\right]\right] \\
&= \frac{1}{ \PP(\bm{X} \in w)} \E\left[\frac{I(\bm{X} \in w, A = a)}{\PP(A =a|\bm{X})}Y \right] 
\end{align*}
\QEDB

\subsection{Proof of Theorem \ref{thm: ipw-var}} \label{app: ipw_var}

\textbf{Proof of Theorem \ref{thm: ipw-var}:} We derive the asymptotic variance when the propensity scores are estimated by fitting a correctly specified logistic regression model using the data falling in the union of the subgroups $p=\{l, r\}$. For simplicity of notation, we denote $e = e_1(\bm{X}; \bm{\beta}) = \{1 + \exp(-\bm{X} \bm{\beta})\}^{-1}$ and $1-e = e_0(\bm{X}; \bm{\beta})$, $\bm{e}_{\bm{\beta}} = \partial e/\partial \bm{\beta}$, $p_l = \PP(\bm{X} \in l | \bm{X} \in p)$. Define $\bm{\theta} = (T(l, r), \bm{\beta}^T, p_l)^T$ as the vector of unknown parameters with the true value denoted as $\bm{\theta}_0$. The joint set of estimating equations used to estimate $\bm{\theta}_0$ using the data falling in the union of the subgroups $p$ is given by
\begin{align*}
    \begin{cases}
    \frac{1}{n(p)}\sum_{i=1}^{n(p)} \psi_1(Y_i, A_i, \bm{X}_i, T(l, r), \bm{\beta}, p_l) = \frac{1}{n(p)}\sum_{i=1}^{n(p)} \left\{ \frac{I(\bm{X}_i \in l)}{p_l} \left[ \frac{A_iY_i}{e_i} - \frac{ (1-A_i) Y_i }{1 - e_i} \right] - \frac{I(\bm{X}_i \in r)}{1-p_l} \left[ \frac{A_iY_i}{e_i}  \right.\right.\\
    \qquad \qquad \qquad \qquad \qquad \qquad \qquad \qquad \qquad \qquad \qquad - \left.\left. \frac{ (1-A_i) Y_i }{1 - e_i} \right]-  T(l, r) \right\} = 0\\
    \frac{1}{n(p)}\sum_{i=1}^{n(p)} \bm{\psi}_2(Y_i, A_i, \bm{X}_i, T(l, r), \bm{\beta}, p_l) = \frac{1}{n(p)}\sum_{i=1}^{n(p)} \frac{(A_i - e_i) \bm{e}_{\bm{\beta}i}}{e_i (1-e_i)} = 0\\
    \frac{1}{n(p)}\sum_{i=1}^{n(p)} \psi_3(Y_i, A_i, \bm{X}_i, T(l, r), \bm{\beta}, p_l) = \frac{1}{n(p)}\sum_{i=1}^{n(p)} \left[ I(\bm{X}_i \in l) - p_l \right] = 0,
    \end{cases} 
\end{align*}
where $\sum_{i=1}^{n(p)}$ refers to summing over all elements in the union of the subgroups $p$. Define $\bm{\hat \theta}_0$ as the solution to the three estimating equations listed above and let $\bm{O} = (Y, A, \bm{X})$ and $\bm{\psi} = (\psi_1, \bm{\psi}_2, \psi_3)$. Here, $\bm{\psi}_2$ is the estimating equation corresponding to the maximum likelihdood estimation from the logistic regression model. Straightforward calculations give
\begin{align*}
    \mathbf{A}(\bm{\theta}_0) =& \E \left[ - \frac{\partial}{\partial \theta} \bm{\psi}(\bm{O}, \bm{\theta}_0) \right]= \begin{pmatrix}
    1 & \bm{H}_{\bm{\beta}, l}^T - \bm{H}_{\bm{\beta}, r}^T & \frac{1}{p_l} [\mu_1(l) - \mu_0(l)] + \frac{1}{p_r}  [\mu_1(r) - \mu_0(r)]\\
    \bm{0} & \mathbf{E}_{\bm{\beta}\bm{\beta}} & \bm{0}\\
    0 & \bm{0} & 1
    \end{pmatrix}\\
    \left[\mathbf{A}(\bm{\theta}_0)\right]^{-1} =& \begin{pmatrix}
    1 & - (\bm{H}_{\bm{\beta}, l} - \bm{H}_{\bm{\beta}, r})^T\mathbf{E}_{\bm{\beta}\bm{\beta}}^{-1} & -\frac{1}{p_l} [\mu_1(l) - \mu_0(l)] - \frac{1}{p_r} [\mu_1(r) - \mu_0(r)]\\
    \bm{0} & \mathbf{E}_{\bm{\beta}\bm{\beta}}^{-1} & \bm{0}\\
    0 & \bm{0} & 1
    \end{pmatrix},
\end{align*}
where as in main manuscript $\bm{H}_{\bm{\beta}, s} =\left. \E\left[ \left(\frac{A Y}{e} + \frac{(1-A) Y}{1-e}\right) \bm{e}_{\bm{\beta}} \bigg|\bm{X} \in s\right]\right|_{\bm{\theta} = \bm{\theta}_{0}}$, for $s = l, r$, and $\mathbf{E}_{\bm{\beta}\bm{\beta}} = \E \left.\left[ \left. \frac{\bm{e}_{\bm{\beta}} \bm{e}_{\bm{\beta}}^T}{e (1- e)} \right| \bm{X} \in p\right] \right|_{\bm{\theta} = \bm{\theta}_{0}}$. Also, 
\begin{align*}
    \mathbf{B}(\bm{\theta}_0) =& \E \left[ \bm{\psi}(\bm{O}, \bm{\theta}_0) \bm{\psi}(\bm{O}, \bm{\theta}_0)^T \right]\\
    =& \begin{pmatrix}
    \left[\mathbf{B}(\bm{\theta}_{0})\right]_{1,1} & (\bm{H}_{\bm{\beta}, l} - \bm{H}_{\bm{\beta}, r})^T & \left[\mathbf{B}(\bm{\theta}_{0})\right]_{1,3}\\
    \bm{H}_{\bm{\beta}, l} - \bm{H}_{\bm{\beta}, r} & \mathbf{E}_{\bm{\beta}\bm{\beta}} & \bm{0}\\
    \left[\mathbf{B}(\bm{\theta}_{0})\right]_{3,1} & \bm{0} & p_l (1-p_l)\\
    \end{pmatrix},
\end{align*}
where $\small \left[\mathbf{B}(\bm{\theta}_{0})\right]_{1,1} = \frac{1}{p_l} \left. \E\left(\left. \frac{\left[AY\right]^2 }{e} + \frac{\left[(1-A)Y\right]^2 }{ 1-e} \right| \bm{X} \in l\right) \right|_{\bm{\theta} = \bm{\theta}_0}+ \frac{1}{p_r} \left. \E\left(\left. \frac{\left[AY\right]^2 }{e} + \frac{\left[(1-A)Y\right]^2 }{ 1-e } \right| \bm{X} \in r\right)\right|_{\bm{\theta} = \bm{\theta}_0} - T(l,r)^2$ and $\left[\mathbf{B}(\bm{\theta}_{0})\right]_{1,3} = \left[\mathbf{B}(\bm{\theta}_{0})\right]_{3,1} = p_r [\mu_1(l) - \mu_0(l)] + p_l [\mu_1(r) - \mu_0(r)]$.

Results in \cite{stefanski2002calculus} imply that the asymptotic variance of $\bm{\hat \theta}_0$ is given by
\begin{align*}
    \mathbf{V}(\bm{\theta}_0) =& \left[\mathbf{A}(\bm{\theta}_0)\right]^{-1}\mathbf{B}(\bm{\theta}_0) \left[ \mathbf{A}(\bm{\theta}_0)^{-1} \right]^T = \begin{pmatrix}
    \left[\mathbf{V}(\bm{\theta}_0)\right]_{1,1} & \bm{0}^T & 0\\
    \bm{0} & \mathbf{E}_{\bm{\beta}\bm{\beta}}^{-1} & \bm{0}\\
    0 & \bm{0} & p_l (1-p_l)
    \end{pmatrix}.
\end{align*}
where $\small \left[\mathbf{V}(\bm{\theta}_0)\right]_{1,1} = \left[\mathbf{B}(\bm{\theta}_{0})\right]_{1,1} - (\bm{H}_{\bm{\beta}, l} - \bm{H}_{\bm{\beta}, r})^T \mathbf{E}_{\bm{\beta}\bm{\beta}}^{-1} (\bm{H}_{\bm{\beta}, l} - \bm{H}_{\bm{\beta}, r}) - \frac{1}{p_lp_r} \left(p_r [\mu_1(l) - \mu_0(l)] + p_l [\mu_1(r) - \mu_0(r)]\right)^2$. $\left[\mathbf{V}(\bm{\theta}_0)\right]_{1,1}$ gives the asymptotic variance of $\hat{T}_{\text{IPW}}(l, r)$. 
\QEDB

\begin{thm} \label{thm: ipw_var_est}
Assume that the propensity scores are estimated by fitting a correctly specified logistic regression model using the data falling in the union of the subgroups $p = \{l,r\}$. The asymptotic variance of $\hat T_{IPW}(l,r)$ when the union of the subgroups $p = \{l, r\}$ is split into $l$ and $r$ can be consistently estimated by
$$\reallywidehat{\text{Var}}\left[\hat{T}_{\text{IPW}}(l, r) \right] =\frac{1}{n(p)}\left(\frac{1}{n(p)}\sum_{i = 1}^{n} I(\bm{X}_i \in p) \hat{I}^2_i - \frac{1}{\hat{p}_l \hat{p}_r} \left[\hat{p}_r (\hat{\mu}_1(l) - \hat{\mu}_0(l)) + \hat{p}_l  (\hat{\mu}_1(r) - \hat{\mu}_0(r)) \right]^2\right),$$ 
where
\begin{align*}
    \hat{I}_i =& \left[ \frac{I(\bm{X}_i \in l)A_i Y_i}{ \hat{p}_l \cdot \hat{e}_i} - \frac{I(\bm{X}_i \in l) (1-A_i) Y_i}{ \hat{p}_l \cdot (1 -\hat{e}_i)} \right] - \left[ \frac{I(\bm{X}_i \in r)A_i Y_i}{ \hat{p}_r \cdot \hat{e}_i} - \frac{I(\bm{X}_i \in r) (1-A_i) Y_i}{ \hat{p}_r \cdot (1 -\hat{e}_i)} \right] \\
    &- \hat{T}_{\text{IPW}}(l, r) - (A_i - \hat{e}_i) \left(\hat{\bm{H}}_{\bm{\beta}, l} - \hat{\bm{H}}_{\bm{\beta}, r} \right)^T \hat{\mathbf{E}}_{\bm{\beta}\bm{\beta}}^{-1} \bm{X}_i
\end{align*}
and
\begin{align*}
    \hat{p}_s =& \frac{n(s)}{n(p)}, \qquad \hat{e}_i = e_1(\bm{X}_i; \hat{\bm{\beta}}_{\text{LR}}) \\
    \hat{\mathbf{E}}_{\bm{\beta}\bm{\beta}} =& \frac{1}{n(p)} \sum_{i = 1}^{n} I(\bm{X}_i \in p) \hat{e}_i (1 -\hat{e}_i)\bm{X}_i\bm{X}_i^T\\
    \hat{\bm{H}}_{\bm{\beta}, s} =& \frac{1}{n(p)} \sum_{i = 1}^{n} \left[ \frac{A_i Y_i (1 -\hat{e}_i) }{\hat{e}_i } + \frac{(1-A_i) Y_i \hat{e}_i }{(1-\hat{e}_i) } \right]\frac{\bm{X}_i I(\bm{X}_i \in s)}{\hat{p}_s}
\end{align*}
for $s = l, r$.
\end{thm}

\textbf{Proof of Theorem \ref{thm: ipw_var_est}:} By the consistency of $\hat{\mu}_a(s), a \in \{0,1\}, s \in \{l,r\}$, $\hat p_l$, and $\hat p_r$ we have 
\[
\frac{1}{\hat{p}_l \hat{p}_r} \left[\hat{p}_r (\hat{\mu}_1(l) - \hat{\mu}_0(l)) + \hat{p}_l  (\hat{\mu}_1(r) - \hat{\mu}_0(r)) \right]^2 \rightarrow \frac{1}{p_l p_r} \left[p_r (\mu_1(l) - \mu_0(l)) + p_l  (\mu_1(r) - \mu_0(r)) \right]^2
\]
in probability. Also, the empirical mean of the squares of each of the first 3 terms of $\hat{I}_i$ converges in probability to $\left[\mathbf{B}(\bm{\theta}_{0})\right]_{1,1}$. 

The square of the last term in $\hat{I}_i$ and cross product terms when squaring $\hat I_i$ converge to
\begin{align*} 
    & \E\left[ (A - \hat{e})^2 \left(\hat{\bm{H}}_{\bm{\beta}, l} - \hat{\bm{H}}_{\bm{\beta}, r} \right)^T \hat{\mathbf{E}}_{\bm{\beta}\bm{\beta}}^{-1} \bm{X} \bm{X}^T \hat{\mathbf{E}}_{\bm{\beta}\bm{\beta}}^{-1} \left(\hat{\bm{H}}_{\bm{\beta}, l} - \hat{\bm{H}}_{\bm{\beta}, r} \right) \right] \\
    & - 2 \E\left\{ (A - \hat{e}) \left(\hat{\bm{H}}_{\bm{\beta}, l} - \hat{\bm{H}}_{\bm{\beta}, r} \right)^T \hat{\mathbf{E}}_{\bm{\beta}\bm{\beta}}^{-1} \bm{X} \left( \left[ \frac{I(\bm{X} \in l)A Y}{ \frac{n(l)}{n(p)} \cdot \hat{e}} - \frac{I(\bm{X} \in l) (1-A) Y}{ \frac{n(l)}{n(p)} \cdot (1 -\hat{e})} \right] \right. \right.\\
    & - \left.\left. \left[ \frac{I(\bm{X} \in r)A Y}{ \frac{n(r)}{n(p)} \cdot \hat{e}} - \frac{I(\bm{X} \in r) (1-A) Y}{ \frac{n(r)}{n(p)} \cdot (1 -\hat{e})} \right] - \hat{T}_{\text{IPW}}(l, r) \right) \right\}\\
    =& \left(\bm{H}_{\bm{\beta}, l} - \bm{H}_{\bm{\beta}, r} \right)^T \mathbf{E}_{\bm{\beta}\bm{\beta}}^{-1}  \E\left[ (A - \hat{e})^2 \bm{X} \bm{X}^T \right]  \mathbf{E}_{\bm{\beta}\bm{\beta}}^{-1} \left(\bm{H}_{\bm{\beta}, l} - \bm{H}_{\bm{\beta}, r} \right) - 2\left(\bm{H}_{\bm{\beta}, l} - \bm{H}_{\bm{\beta}, r} \right)^T \mathbf{E}_{\bm{\beta}\bm{\beta}}^{-1} \\
    & \cdot \E\left\{ \frac{\bm{X}I(\bm{X} \in l)}{ n(l)/n(p) } \left[ \frac{A Y (1-\hat{e})}{ \hat{e}} + \frac{ (1-A) Y \hat{e} }{ 1 -\hat{e}} \right] - \frac{\bm{X}I(\bm{X} \in r)}{ n(r)/n(p) } \left[ \frac{A Y (1-\hat{e})}{ \hat{e}} + \frac{ (1-A) Y \hat{e} }{ 1 -\hat{e}} \right] \right.\\
    & \left. - (A - \hat{e})\bm{X}\hat{T}_{\text{IPW}}(l, r) \right\}\\
    =& - (\bm{H}_{\bm{\beta}, l} - \bm{H}_{\bm{\beta}, r})^T \mathbf{E}_{\bm{\beta}\bm{\beta}}^{-1} (\bm{H}_{\bm{\beta}, l} - \bm{H}_{\bm{\beta}, r})
\end{align*}
Therefore, $\lim_{n \rightarrow\infty} \frac{1}{n(p)}\sum_{i = 1}^{n} I(\bm{X}_i \in p) \hat{I}^2_i = \left[\mathbf{B}(\bm{\theta}_{0})\right]_{1,1} - (\bm{H}_{\bm{\beta}, l} - \bm{H}_{\bm{\beta}, r})^T \mathbf{E}_{\bm{\beta}\bm{\beta}}^{-1} (\bm{H}_{\bm{\beta}, l} - \bm{H}_{\bm{\beta}, r})$, which completes the proof. 
\QEDB

\subsection{Deriving asymptotic variance of $\hat{T}_{\text{IPW}}(l, r)$ and a consistent variance estimator when propensity scores are estimated separately in the two subgroups} \label{app: ipw_var_insplt}

Implementation of the Causal Interaction Tree algorithms requires choosing what part of the data is used to fit the propensity score model (e.g.,~fit a single model in the union of the subgroups $\{l,r\}$ or fit a separate model in the subgroups $l$ and $r$). Fitting a propensity score model using all the data in $\{l,r\}$ results in a fitted model using a larger sample size compared to fitting a separate model in each of the subgroups, but using a separate model for each of the subgroups is more flexible as it allows the relationship between the treatment and the covariates to differ between the two subgroups. Another alternative that we explore in the simulations is to build a single model before the start of the tree building process and use that model to estimate the propensity scores in all the tree building process. 

In the following two theorems we derive the asymptotic variance and provide a consistent variance estimator for the case when the propensity score model for the inverse probability weighting estimator is fit using a separate logistic regression models in each subgroup.

\begin{thm}
\label{thm: ipw-var-insplt}
Assume that the propensity scores are estimated by fitting two separate correctly specified logistic regression models in the two subgroups $l$ and $r$. The asymptotic variance of $\hat T_{\text{IPW}}(l,r)$ when $p= \{l, r\}$ is split into two subgroups $l$ and $r$ is given by
\begin{align}
    &\frac{1}{\PP(\bm{X} \in l|\bm{X} \in p)} \left. \E\left( \frac{\left[Y A\right]^2 }{e_1(\bm{X}; \bm{\beta}_{\text{LR}, l}) } + \frac{\left[Y (1-A)\right]^2 }{e_0(\bm{X}; \bm{\beta}_{\text{LR}, l}) } \right| \bm{X} \in l\right) \nonumber \\
    &+ \frac{1}{\PP(\bm{X} \in r|\bm{X} \in p)} \left. \E\left( \frac{\left[Y A \right]^2 }{e_1(\bm{X}; \bm{\beta}_{\text{LR}, r}) } + \frac{\left[Y (1 - A)\right]^2 }{ e_0(\bm{X}; \bm{\beta}_{\text{LR}, r}) } \right| \bm{X} \in r\right) - T(l,r)^2 \nonumber  \\ 
    &- \bm{H}_{\bm{\beta}_{\text{LR},l}}^T \mathbf{E}_{\bm{\beta}\bm{\beta}, l}^{-1} \bm{H}_{\bm{\beta}_{\text{LR}, l}} - \bm{H}_{\bm{\beta}_{\text{LR},r}}^T \mathbf{E}_{\bm{\beta}\bm{\beta},r}^{-1} \bm{H}_{\bm{\beta}_{\text{LR},r}} \nonumber \\ 
    &- \frac{\left(\PP(\bm{X} \in r|\bm{X} \in p)\left[\mu_{1}(l) - \mu_{0}(l)\right] + \PP(\bm{X} \in l|\bm{X} \in p) \left[\mu_{1}(r) - \mu_{0}(r)\right] \right)^2}{\PP(\bm{X} \in l|\bm{X} \in p) \PP(\bm{X} \in r|\bm{X} \in p)} \label{eq:var-ipw-insplt}.
\end{align}
Here, for subgroup $s \in \{l,r\}$ $\bm{\beta}_{\text{LR},s}$ is the limit of the maximum likelihood logistic regression estimator and 
\begin{align*} 
\bm{H}_{\bm{\beta}_{\text{LR}, s}} =& \E\left[ \left(\frac{A Y}{e_1(\bm{X}; \bm{\beta}_{\text{LR}, s}) } + \frac{(1-A) Y}{e_0(\bm{X}; \bm{\beta}_{\text{LR}, s}) }\right) \frac{\partial}{\partial \bm{\beta}} e_1(\bm{X}; \bm{\beta}) \bigg |_{\bm{\beta} = \bm{\beta}_{\text{LR}, s}}\bigg|\bm{X} \in s\right] \\ 
\mathbf{E}_{\bm{\beta}\bm{\beta},s} =& \E \left[ \left. \frac{ I (\bm{X} \in s)\frac{ \partial}{\partial \bm{\beta}} e_1(\bm{X}; \bm{\beta}) \bigg |_{\beta = \beta_{LR, s}} \left(\frac{\partial}{\partial \bm{\beta}} e_1(\bm{X}; \bm{\beta}) \bigg |_{\beta = \beta_{LR, s}} \right)^T}{ e_1(\bm{X}; \bm{\beta}_{\text{LR}, s}) e_0(\bm{X}; \bm{\beta}_{\text{LR}, s})} \right| \bm{X} \in s\right].
\end{align*}
\end{thm}
\textbf{Proof of Theorem \ref{thm: ipw-var-insplt}:} 
For simplicity of notation define $e_{i,s} = e_1(\bm{X}_i; \bm{\beta}_{\text{LR}, s}) = \{1 + \exp(-\bm{X}_i \bm{\beta}_{LR,s})\}^{-1}$ and $1 - e_{i,s} = e_0(\bm{X}_i; \bm{\beta}_{\text{LR}, s})$, and define $\bm{e}_{\bm{\beta}_{i, s}} = \partial e_{i,s}/\partial \bm{\beta}_{LR,s}$ for $s \in \{l, r\}$.
The joint set of  estimating equations using the data in the union of the subgroups $p$ for $\bm{\theta} = (T(l, r), \bm{\beta}^T_{LR,l}, \bm{\beta}^T_{LR,r}, p_l)^T$ is given by
\begin{align*}
    \begin{cases}
    \frac{1}{n(p)}\sum_{i=1}^{n(p)} \psi_1(Y_i, A_i, \bm{X}_i, \bm{\theta}) = \frac{1}{n(p)}\sum_{i=1}^{n(p)} \left\{ \frac{I(\bm{X}_i \in l)}{p_l} \left[ \frac{A_iY_i}{e_{i, l}} - \frac{ (1-A_i) Y_i }{1 - e_{i,l}} \right] - \frac{I(\bm{X}_i \in r)}{1-p_l} \left[ \frac{A_iY_i}{e_{i,r}} - \frac{ (1-A_i) Y_i }{1 - e_{i,r}} \right]\right.\\
    \qquad \qquad \qquad \qquad \qquad \qquad \qquad \qquad \left.- T(l,r) \right\} = 0\\
    \frac{1}{n(p)}\sum_{i=1}^{n(p)} \bm{\psi}_2(Y_i, A_i, \bm{X}_i, \bm{\theta}) = \frac{1}{n(p)}\sum_{i=1}^{n(p)} \frac{I(\bm{X}_i\in l)(A_i - e_{i,l}) \bm{e}_{\bm{\beta}i,l}}{e_{i,l} (1-e_{i,l})} = 0\\
    \frac{1}{n(p)}\sum_{i=1}^{n(p)} \bm{\psi}_3(Y_i, A_i, \bm{X}_i, \bm{\theta}) = \frac{1}{n(p)}\sum_{i=1}^{n(p)} \frac{I(\bm{X}_i\in r)(A_i - e_{i,r}) \bm{e}_{\bm{\beta}i,r}}{e_{i,r} (1-e_{i,r})} = 0\\
    \frac{1}{n(p)}\sum_{i=1}^{n(p)} \psi_4(Y_i, A_i, \bm{X}_i, \bm{\theta}) = \frac{1}{n(p)}\sum_{i=1}^{n(p)} \left[ I(\bm{X}_i \in l) - p_l \right] = 0,
    \end{cases}
\end{align*}
where in the above $\sum_{i=1}^{n(p)}$ refers to summing over all elements in $p$. Let $\bm{\theta}_0$ be the true value of the set of parameters $\bm{\theta}$, $\bm{O} = (Y,A,\bm{X})$ and define $\bm{\psi} = (\psi_1, \bm{\psi}_2,\bm{\psi}_3, \psi_4)$.

Analogous calculations to those in the proof of Theorem \ref{thm: ipw-var} show that 
\begin{align*}
    \mathbf{A}(\bm{\theta}_0) =& \E \left[ - \frac{\partial}{\partial \theta} \bm{\psi}(\bm{O}, \bm{\theta}_0) \right]= \begin{pmatrix}
    1 & \bm{H}_{\bm{\beta}_{LR, l}}^T & - \bm{H}_{\bm{\beta}_{LR, r}}^T & \frac{1}{p_l} [\mu_1(l) - \mu_0(l)] + \frac{1}{p_r} [\mu_1(r) - \mu_0(r)]\\
    \bm{0} & \mathbf{E}_{\bm{\beta}\bm{\beta},l} & \bm{0} & \bm{0}\\
    \bm{0} & \bm{0} &\mathbf{E}_{\bm{\beta}\bm{\beta},r} & \bm{0}\\
    0 & \bm{0} & \bm{0} & 1
    \end{pmatrix}\\
    \mathbf{A}^{-1}(\bm{\theta}_0) =& \begin{pmatrix}
    1 & -\bm{H}_{\bm{\beta}_{LR, l}}^T\mathbf{E}_{\bm{\beta}\bm{\beta},l}^{-1} & \bm{H}_{\bm{\beta}_{LR, r}}^T\mathbf{E}_{\bm{\beta}\bm{\beta},r}^{-1} & -\frac{1}{p_l} [\mu_1(l) - \mu_0(l)]- \frac{1}{p_r} [\mu_1(r) - \mu_0(r)] \\
    \bm{0} & \mathbf{E}_{\bm{\beta}\bm{\beta}, l}^{-1} & \bm{0} & \bm{0}\\
    \bm{0} & \bm{0} & \mathbf{E}_{\bm{\beta}\bm{\beta}, r}^{-1} & \bm{0}\\
    0 & \bm{0} & \bm{0} & 1
    \end{pmatrix}
\end{align*}
We have
\begin{align*}
    \mathbf{B}(\bm{\theta}_0) =& \E \left[ \bm{\psi}(\bm{O}, \bm{\theta}_0) \bm{\psi}(\bm{O}, \bm{\theta}_0)^T \right]\\
    =& \begin{pmatrix}
    \left[\mathbf{B}(\bm{\theta}_{0})\right]_{1,1} & \bm{H}_{\bm{\beta}_{LR, l}}^T & - \bm{H}_{\bm{\beta}_{LR, r}}^T & \left[\mathbf{B}(\bm{\theta}_{0})\right]_{1,4} \\
    \bm{H}_{\bm{\beta}_{LR, l}} & \mathbf{E}_{\bm{\beta}\bm{\beta},l } & \bm{0} & \bm{0}\\
    -\bm{H}_{\bm{\beta}_{LR, r}} &\bm{0} & \mathbf{E}_{\bm{\beta}\bm{\beta},r} & \bm{0}\\
    \left[\mathbf{B}(\bm{\theta}_{0})\right]_{4,1} & \bm{0} & \bm{0} & p_l (1-p_l)\\
    \end{pmatrix}
\end{align*}
where $\left[\mathbf{B}(\bm{\theta}_{0})\right]_{1,1} = \left.\E\left[\psi_1^2(Y_i, A_i, \bm{X}_i, \bm{\theta})\right]\right|_{\bm{\theta} = \bm{\theta}_0}$ and $\left[\mathbf{B}(\bm{\theta}_{0})\right]_{1,4} = \left[\mathbf{B}(\bm{\theta}_{0})\right]_{4,1} = p_r [\mu_1(l) - \mu_0(l)] + p_l [\mu_1(r) - \mu_0(r)]$. Results in \cite{stefanski2002calculus} imply that the asymptotic variance of $\bm{\hat \theta}_0$ is given by
    $\mathbf{V}(\bm{\theta}_0) = \left[\mathbf{A}(\bm{\theta}_0)\right]^{-1}\mathbf{B}(\bm{\theta}_0) \left[ \mathbf{A}(\bm{\theta}_0)^{-1} \right]^T.$ 
It follows that the asymptotic variance of $\hat{T}_{\text{IPW}}(l, r)$ is
\begin{align*}
     \left[\mathbf{B}(\bm{\theta}_{0})\right]_{1,1} - \bm{H}_{\bm{\beta}, l}^T \mathbf{E}_{\bm{\beta}\bm{\beta}, l}^{-1} \bm{H}_{\bm{\beta}, l}  - \bm{H}_{\bm{\beta}, r}^T \mathbf{E}_{\bm{\beta}\bm{\beta}, r}^{-1} \bm{H}_{\bm{\beta}, r} - \frac{1}{p_lp_r} \left(p_r [\mu_1(l) - \mu_0(l)] + p_l [\mu_1(r) - \mu_0(r)]\right)^2.
\end{align*}
\QEDB

From the above theorem we see that the variance estimator can be decomposed into four parts: one corresponding to the variance when the true propensity scores and subgroup probabilities are used to calculate $\hat T_{\text{IPW}}(l, r)$ and three terms corresponding to the variance reduction associated with estimating each of the two propensity score models and the subgroup probability.

\begin{thm} \label{thm: ipw-var-insplt-est}
Assume that the propensity scores are estimated by fitting two separate correctly specified logistic regression model in the two subgroups $l$ and $r$. The asymptotic variance of $\hat T_{\text{IPW}}(l,r)$ when the union of the subgroups $p = \{l, r\}$ is split into two subgroups $l$ and $r$ can be consistently estimated by
$$\reallywidehat{\text{Var}}\left[\hat{T}_{\text{IPW}}(l,r)\right] =\frac{1}{n(p)}\left(\frac{1}{n(p)}\sum_{i = 1}^{n} I(\bm{X}_i \in p) \hat{I}^2_i - \frac{1}{\hat{p}_l \hat{p}_r} \left[\hat{p}_r (\hat{\mu}_1(l) - \hat{\mu}_0(l)) + \hat{p}_l  (\hat{\mu}_1(r) - \hat{\mu}_0(r)) \right]^2 \right)$$
where
\begin{align*}
    \hat{I}_i =& \left[ \frac{I(\bm{X}_i \in l)A_i Y_i}{ \hat{p}_l \cdot \hat{e}_{i,l}} - \frac{I(\bm{X}_i \in l) (1-A_i) Y_i}{ \hat{p}_l \cdot (1 -\hat{e}_{i,l})} \right] - \left[ \frac{I(\bm{X}_i \in r)A_i Y_i}{ \hat{p}_r \cdot \hat{e}_{i,r}} - \frac{I(\bm{X}_i \in r) (1-A_i) Y_i}{ \hat{p}_r \cdot (1 -\hat{e}_{i,l})} \right] \\
    & - \hat{T}_{\text{IPW}}(l,r) - I(\bm{X}_i \in l) (A_i - \hat{e}_{i,l}) \hat{\bm{H}}_{\bm{\beta}, l}^T \hat{\mathbf{E}}_{\bm{\beta}\bm{\beta}, l}^{-1} \bm{X}_i - I(\bm{X}_i \in r) (A_i - \hat{e}_{i,r}) \hat{\bm{H}}_{\bm{\beta}, r}^T \hat{\mathbf{E}}_{\bm{\beta}\bm{\beta}, r}^{-1} \bm{X}_i
\end{align*}
and
\begin{align*}
    \hat{p}_s =& \frac{n(s)}{n(p)}, \qquad \hat{e}_{i,s} = e_1(\bm{X}_i; \hat{\bm{\beta}}_{\text{LR}, s}) \\
    \hat{\mathbf{E}}_{\bm{\beta}\bm{\beta},s} =& \frac{1}{n(s)} \sum_{i = 1}^{n} I(\bm{X}_i \in s) \hat{e}_{i,s} (1 -\hat{e}_{i,s})\bm{X}_i\bm{X}_i^T\\
    \hat{\bm{H}}_{\bm{\beta}, s} =& \frac{1}{n(p)} \sum_{i = 1}^{n} \left[ \frac{A_i Y_i (1 -\hat{e}_{i,s}) }{\hat{e}_{i,s} } + \frac{(1-A_i) Y_i \hat{e}_{i,s} }{(1-\hat{e}_{i,s}) } \right]\frac{\bm{X}_i I(\bm{X}_i \in s)}{\hat{p}_s}
\end{align*}
for $s = l, r$.
\end{thm} 

The proof is omitted as it is similar to the proof of Theorem \ref{thm: ipw_var_est}.

\subsection{The first order influence function of $\mu_a(w)$ under non-parametric model} \label{app: IF_DR}

Let $\{p_t: t\in [0, 1)\}$ be a regular parametric submodel with $t=0$ being the true "data law". Let $\bm{O} = (Y, \bm{X}, A)$ be the observed data, and $l(\cdot)$ be the score function. Calculating the pathwise derivative of the target parameter using identification result \eqref{gform-id} and evaluating at $t=0$ gives
\begin{align*}
    & \left.\frac{\partial }{\partial t} \E_{p_t}[\E_{p_t}[Y|\bm{X}, A = a]|\bm{X} \in w] \right|_{t=0} \\
    =& \left.\frac{\partial}{\partial t} \E_{p_t}[\E[Y|\bm{X}, A = a]|\bm{X} \in w] \right|_{t=0} + \E\left[ \left.\frac{\partial}{\partial t} \E_{p_t}[Y|\bm{X}, A = a] \right|_{t=0} |\bm{X} \in w\right]\\
    =& \underbrace{\E\big\{ \E[Y|\bm{X}, A = a] l(\bm{O}|\bm{X} \in w)|\bm{X} \in w\big\}}_{T_1} + \underbrace{\E\big\{  \E[Y l(Y|\bm{X}, A = a)|\bm{X}, A = a]  |\bm{X} \in w\big\}}_{T_2}.
\end{align*}
Since,
\begin{align*}
    \E\{\mu_a(w) l(\bm{O}|\bm{X} \in w) | \bm{X} \in w \} = \mu_a(w) \E\{ l(\bm{O}|\bm{X} \in w) | \bm{X} \in w \} = 0
\end{align*}
we have
\begin{align*}
    T_1 =& \E\left\{ \left[\E(Y|\bm{X}, A = a)  - \mu_a(w) \right] l(\bm{O}|\bm{X} \in w)|\bm{X} \in w\right\} \\ 
    =& \frac{1}{\PP(\bm{X} \in w)} \E\left\{ I(\bm{X} \in w) \left[\E(Y|\bm{X}, A = a)  - \mu_a(w) \right] l(\bm{O}|\bm{X} \in w)\right\} \\
    =& \frac{1}{\PP(\bm{X} \in w)} \E\left\{ I(\bm{X} \in w) \left[\E(Y|\bm{X}, A = a)  - \mu_a(w) \right] \frac{\partial}{\partial t} \log \PP_t(\bm{O}| \bm{X} \in w)\right\}\\
    =& \frac{1}{\PP(\bm{X} \in w)} \E\left\{ I(\bm{X} \in w) \left[\E(Y|\bm{X}, A = a)  - \mu_a(w) \right] l(\bm{O}) \right\}\\
    & - \underbrace{\frac{1}{\PP(\bm{X} \in w)} \E\left\{ I(\bm{X} \in w) \left[\E(Y|\bm{X}, A = a) - \mu_a(w) \right] \frac{\partial}{\partial t} \log \PP_t(\bm{X} \in w)\right\}}_{T_3}
\end{align*}
\begin{align*}
    T_3 =& \frac{\partial}{\partial t} \log \PP_t(\bm{X} \in w) \E\left\{ \E(Y|\bm{X}, A = a) - \mu_a(w)  | \bm{X} \in w \right\} \\
    =& \frac{\partial}{\partial t} \log \PP_t(\bm{X} \in w) \left[\E\left\{ \E(Y|\bm{X}, A = a) | \bm{X} \in w \right\} - \mu_a(w)\right] = 0.
\end{align*}
Using that 
\begin{align*}
    & \E\left[  \E[\E(Y| \bm{X}, A = a)l(Y|\bm{X}, A = a)|\bm{X}, A = a]  |\bm{X} \in w\right]\\
    =&  \E\left[  \E(Y| \bm{X}, A = a) \E[l(Y|\bm{X}, A = a)|\bm{X}, A = a]  |\bm{X} \in w\right] = 0,
\end{align*}
we get
\begin{align*}
    T_2=& \E\left\{ \left.\E\left[\left[Y - \E(Y| \bm{X}, A = a) \right] l(Y|\bm{X}, A = a) |\bm{X}, A = a\right]  \right|\bm{X} \in w\right\} \\
    =& \E\left\{ \left. \E\left[ \left.\frac{I(A = a)}{\PP(A=a|\bm{X})} \left[Y - \E(Y| \bm{X}, A = a) \right] l(Y|\bm{X}, A = a) \right|\bm{X} \right]  \right|\bm{X} \in w\right\} \\
    =& \E\left\{  \left. \E\left[\left.\frac{I(A = a)}{\PP(A=a|\bm{X})} \left[Y - \E(Y| \bm{X}, A = a) \right] l(Y|\bm{X}, A) \right|\bm{X} \right] \right|\bm{X} \in w\right\} \\
    =& \frac{1}{\PP(\bm{X} \in w)} \E\left\{ \frac{I(\bm{X} \in w) I(A = a)}{\PP(A=a|\bm{X})} \left[Y - \E(Y| \bm{X}, A = a) \right] l(Y|\bm{X}, A)  \right\} \\
    =& \frac{1}{\PP(\bm{X} \in w)} \E\left\{ \frac{I(\bm{X} \in w) I(A = a)}{\PP(A=a|\bm{X})} \left[Y - \E(Y| \bm{X}, A = a) \right] l(\bm{O})  \right\} \\
    & - \underbrace{ \frac{1}{\PP(\bm{X} \in w)}  \E\left\{ \frac{I(\bm{X} \in w) I(A = a)}{\PP(A=a|\bm{X})} \left[Y - \E(Y| \bm{X}, A = a) \right] l(\bm{X}, A)  \right\}}_{T_4}
\end{align*}
Now
\begin{align*}
    T_4 =& \E\left\{ \left. \E\left[ \left. \frac{ I(A = a) }{ \PP(A=a|\bm{X}) } \left[Y - \E(Y| \bm{X}, A = a) \right] l(\bm{X}, A=a) \right| \bm{X}\right] \right|\bm{X} \in w\right\}\\
    =& 0
\end{align*}
Combing the previous results gives, 
\begin{align*}
    \left.\frac{\partial}{\partial t} \E_{p_t}[\E_{p_t}[Y|\bm{X}, A = a]|\bm{X} \in w] \right|_{t=0} =& \E\left\{ \frac{1}{\PP(\bm{X} \in w)} \left(I(\bm{X} \in w) \left[\E(Y|\bm{X}, A = a)  - \mu_a(w)\right] \right.\right.\\
    &\quad \left.\left. + \frac{I(\bm{X} \in w) I(A = a)}{\PP(A=a|\bm{X})} \left[Y - \E(Y| \bm{X}, A = a) \right] \right) l(\bm{O}) \right\}
\end{align*}
which completes the derivation of the influence function.

\subsection{Consistency and asymptotic properties of the doubly robust estimator} \label{app: DR_Inference}

\textbf{Proof of Theorem \ref{Thm-DR}:} 

Unless otherwise stated, all convergence results in this section refer to convergence in probability.
\begin{enumerate}
    \item \textit{Consistency:} 
    By the law of large numbers,
\begin{align*}
     \hat{\mu}_{\text{DR}, a}(w) \rightarrow&   \E\left[\frac{I(\bm{X}\in w) }{\PP(\bm{X}\in w)}\left( g_a(\bm{X}; \bm{\eta}^*_a) + \frac{I(A = a)}{e_a(\bm{X}; \bm{\beta}^*)} \left[Y - g_a(\bm{X}; \bm{\eta}^*_a) \right]\right)\right]
\end{align*}
We now study the asymptotic limit of the doubly robust estimator by considering the two different cases in Assumption A.4.

\begin{itemize}
    \item When $g_a(\bm{X}; \hat{\bm{\eta}}_a)  \rightarrow \E[Y|\bm{X}, A=a]$ holds, using the law of total expectation,
    \begin{align*}
    \hat{\mu}_{\text{DR}, a}(w) \rightarrow& \E\left[ \left. \E[Y|\bm{X}, A=a] + \frac{I(A = a)}{ e_a(\bm{X}; \bm{\beta}^*) } \left[Y - \E(Y|\bm{X}, A=a)\right] \right| \bm{X}\in w \right]\\
    =& \E\left[ \E[Y|\bm{X}, A=a]|\bm{X} \in w\right] \\
    =& \mu_a(w)
\end{align*}
    
    \item When $e_a(\bm{X}; \hat{\bm{\beta}}) \rightarrow P(A=a |\bm{X})$ holds, using the law of total expectation gives
    \begin{align*}
     \hat{\mu}_{\text{DR}, a}(w) \rightarrow& \E\left[ \left. g_a(\bm{X}; \bm{\eta}^*_a) + \frac{I(A = a)}{P(A=a |\bm{X})} \left[Y - g_a(\bm{X}; \bm{\eta}^*_a) \right] \right| \bm{X}\in w \right]\\
     =& \E\left[ \left. g_a(\bm{X}; \bm{\eta}^*_a) + \frac{\E\left[I(A = a)Y | \bm{X}\right] - g_a(\bm{X}; \bm{\eta}^*_a) P(A=a |\bm{X}) }{P(A=a |\bm{X})} \right| \bm{X}\in w \right] \\
     =& \E\left[ \left. \E\left[Y | \bm{X}, A = a\right]  \right| \bm{X}\in w \right] \\
     =& \mu_a(w)
\end{align*}

\end{itemize}
This completes the consistency proof of $\hat{\mu}_{\text{DR},a}(w)$.

\item \textit{Rate of convergence:}

Decompose
\begin{align*}
&\sqrt{n}\left(\hat{\mu}_{\text{DR,a}}(w) - \mu_a(w) \right) \\
=&  \underbrace{\left\{ \mathbb{G}_n\left[H( e_a(\bm{X}; \hat{\bm{\beta}}), g_a(\bm{X}; \hat{\bm{\eta}}_a), \hat{\gamma})\right] -\mathbb{G}_n(H( e_a(\bm{X}; \bm{\beta}^*) , g_a(\bm{X}; \bm{\eta}^*_a), \gamma)) \right\}}_{T_1} \\
& + \underbrace{\mathbb{G}_n(H(e_a(\bm{X}; \bm{\beta}^*), g_a(\bm{X}; \bm{\eta}^*_a), \gamma))}_{T_2} \\
& +\underbrace{\sqrt{n} \left\{ \E   \left[H(e_a(\bm{X}; \hat{\bm{\beta}}), g_a(\bm{X}; \hat{\bm{\eta}}_a), \hat{\gamma})\right] - \mu_a(w) \right\}}_{T_3}
\end{align*}
By assumptions A.1 and A.2, $T_1 = o_P(1)$. By the central limit theorem and assumption A.3, $T_2$ is asymptotically normal and $\sqrt{n}$ consistent. Using that $\sqrt{n}(\hat \gamma - \gamma) = O_P(1)$ we have 

\begin{align*}
    \frac{1}{\sqrt{n}} T_3 =& \E \left\{ \hat{\gamma} I(\bm{X}\in w) \left( g_a(\bm{X}; \hat{\bm{\eta}}_a) + \frac{I(A = a)}{ e_a(\bm{X}; \hat{\bm{\beta}}) } \left(Y - g_a(\bm{X}; \hat{\bm{\eta}}_a) \right) \right) \right\} - \mu_a(w) \\ 
    =& \E \left\{ \gamma I(\bm{X}\in w) \left( g_a(\bm{X}; \hat{\bm{\eta}}_a) + \frac{I(A = a)}{ e_a(\bm{X}; \hat{\bm{\beta}}) } \left(Y - g_a(\bm{X}; \hat{\bm{\eta}}_a) \right) \right) \right\} - \mu_a(w) + O_P\left(\frac{1}{\sqrt{n}}\right).
\end{align*}
Using the law of total expectation, the Cauchy Schwarz inequality, and the identifiability result \eqref{gform-id} we have
\begin{align*}
&\E \left\{ \gamma I(\bm{X}\in w) \left( g_a(\bm{X}; \hat{\bm{\eta}}_a) + \frac{I(A = a)}{ e_a(\bm{X}; \hat{\bm{\beta}}) } \left(Y - g_a(\bm{X}; \hat{\bm{\eta}}_a) \right) \right) \right\} - \mu_a(w)
   \\ &\leq O_P \left( \vphantom{\frac{1}{n}} || e_a(\bm{X}; \hat{\bm{\beta}}) - \PP(A=a |\bm{X}) ||_2 \times || g_a(\bm{X}; \hat{\bm{\eta}}_a) - \E[Y|\bm{X}, A=a] ||_2  \right).
\end{align*}
The rate of convergence follows by combining the above results. 
\end{enumerate}

\section{Alternative Final Tree Selection Method}
\label{app: FTS2}

The alternative final tree selection method is adapted from the tree selection method proposed in \cite{steingrimsson2019subgroup}. We will now briefly describe the method, but refer to \cite{steingrimsson2019subgroup} for further details. In the Classification and Regression Tree algorithm for outcome prediction \citep{breiman1984classification}, the final tree is selected based on minimizing cross-validation error. As the treatment effect is not observed on any participant, cross-validation cannot be directly used for final tree selection using treatment effect prediction error.
    
To overcome this difficulty, we will use the random forest algorithm \citep{breiman2001random} as a surrogate for the true treatment effect estimate and select the tree that gives the prediction closest to the random forest predictions. 
We start by splitting the data into a training and a validation set, fitting an inverse probability weighted random forest model to the training data, and calculating the treatment effect predictions on the validation set. We refer to the predictions as the random forest treatment effect validation set predictions.

In the final tree selection step, for a given split into a training and a validation set and a candidate tree  $\hat{\psi}_m$, $m \in\{1,\dots, \hat{M}\}$, re-estimate the terminal node estimators of $\hat{\psi}_m$ using only the training data falling in each terminal node. Use the re-estimated terminal node estimators to predict the treatment effect for all participants in the validation set and refer to the predictions as the Causal Interaction Tree validation set predictions. Calculate the cross-validation error for tree $\hat{\psi}_m$ corresponding to this particular split into validation and training set as the average $L_2$ distance between the Causal Interaction Tree validation set predictions and the random forest treatment effect validation set predictions. The final tree is selected as the tree that results in the smallest cross-validation error averaged over all splits into validation and training sets.

\section{Additional simulation results} \label{app: add_sim}

We use 1000 simulations to evaluate all the algorithms in this section, except in Appendix \ref{app: cv1_time}, where the running time for the algorithms compared in Section \ref{sec: simulations} is evaluated using 10,000 simulations.

\subsection{Selecting a splitting rule and cross-validation method for Causal Tree algorithms} \label{app: ct_sim}

In this section, we identify the combination of splitting rule and cross-validation method in \texttt{causalTree} that performs the best in terms of MSE in our simulation settings.

The splitting rule (\texttt{split.Rule}) can be chosen from transformed outcome trees (\texttt{TOT}), causal trees (\texttt{CT}), fit-based trees (\texttt{fit}) and squared t-statistic trees (\texttt{tstats}). Both adaptive and honest versions of the estimators are available for the splitting rules except for \texttt{TOT}. This leads to 7 possible choices of splitting rules. The cross-validation method (\texttt{cv.option}) can be chosen from transformed outcome (\texttt{TOT}), matching (\texttt{matching}), \texttt{CT} and \texttt{fit}. Adaptive and honest versions of the criterion are available for \texttt{CT} and \texttt{fit}. This leads to 6 possible choices of cross-validation methods. 

\begin{figure}[htbp]
        \begin{center}
        \centerline{\includegraphics[width = 6.5in]{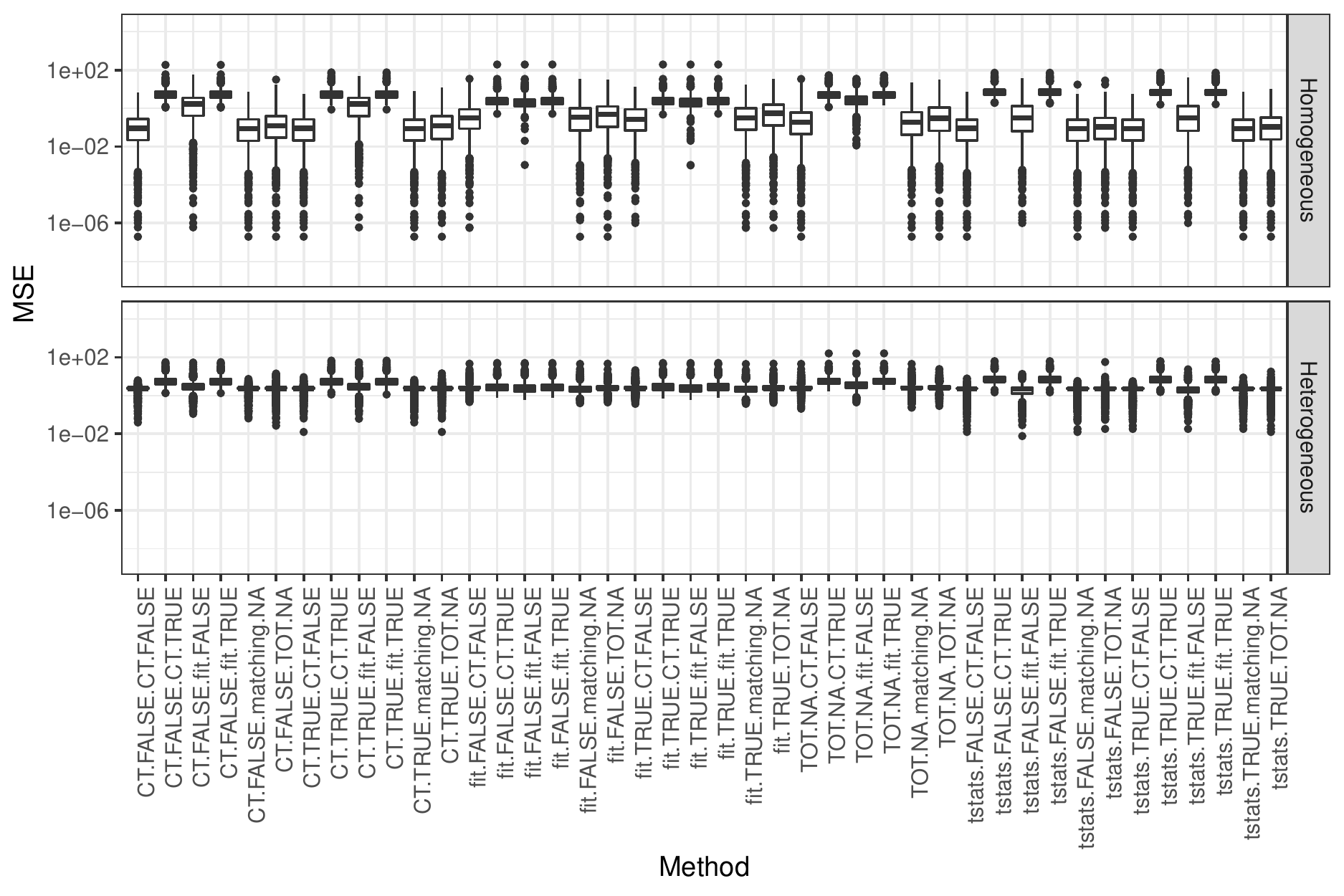}}
    \end{center}
    \caption{Mean squared error (MSE) for different choices of splitting rules and cross-validation methods for the Causal Tree algorithms for the homogeneous (top) and the heterogeneous (bottom) settings described in Section \ref{sec: sim_setup}. Lower values indicate better performance.  Splitting rule (\texttt{split.Rule}) can be chosen from transformed outcome trees (\texttt{TOT}), causal tree (\texttt{CT}), fit-based trees (\texttt{fit}) and squared t-statistic trees (\texttt{tstats}). Adaptive (\texttt{split.Honest = FALSE}) and honest (\texttt{split.Honest = TRUE}) versions are available except for \texttt{TOT}. Cross-validation method (\texttt{cv.option}) can be chosen from transformed outcome \texttt{TOT}, matching (\texttt{matching}), \texttt{CT} and \texttt{fit}. Adaptive (\texttt{cv.Honest = FALSE}) and honest (\texttt{cv.honest = TRUE}) versions are available for \texttt{CT} and \texttt{fit}. The naming pattern for the choices are "(\texttt{split.Rule}).(\texttt{split.Honest}).(\texttt{cv.option}).(\texttt{cv.Honest})". If there is no honest version of the chosen splitting rule or cross-validation method, \texttt{split.Honest} or \texttt{cv.Honest} is \texttt{NA}.}
    \label{fig: ct_settings}
\end{figure}

Figure \ref{fig: ct_settings} shows boxplots of MSE using 1000 simulations for both the homogeneous and the heterogeneous simulation settings described in Section \ref{sec: sim_setup}. All  combinations of splitting rules and cross-validation methods are included in the boxplots (a total of $7 \times 6 = 42$ combinations). The propensity score model adjusting for \texttt{weights} includes the main effects of all covariates. Table \ref{tab: ct_settings} shows the average MSE, proportion of correct trees, average number of noise variables, pairwise prediction similarity, the average running time for both simulation settings, and the proportion of trees making a correct first split in the hetetogeneous setting for the ten combinations of splitting rules and cross-validation methods that have the highest rank on average in terms of minimizing MSE.

The results in Figure \ref{fig: ct_settings} and Table \ref{tab: ct_settings} show that setting \texttt{split.Rule} equal to \texttt{"tstats"} with the honest version (\texttt{split.Honest = TRUE}) and \texttt{cv.option} to \texttt{"matching"} performs the best and is therefore used for the "Best CT" in the main manuscript. 


\begin{table}[htbp]
\centering
\rotatebox{90}{ 
 \begin{minipage}{22cm}
\footnotesize\centering
\begin{tabular}{@{}llccccccccccccc@{}}
\toprule
&\multicolumn{4}{c}{Homogeneous Effect} &  & \multicolumn{6}{c}{Heterogeneous Effect}\\
\cmidrule[0.5pt]{2-6} \cmidrule[0.5pt]{8-13}
 & & Correct & Number & & & & & Correct & Number & & & Correct\\
 Parameter Setting & MSE & Trees & Noise & PPS & Time & & MSE & Trees & Noise & PPS & Time & First Split \\ 
  \hline
tstats TRUE matching NA & 0.23 & 0.98 & 0.04 & 1.00 & 0.02 &  & 2.14 & 0.15 & 0.08 & 0.58 & 0.02 & 0.63 \\ 
  tstats TRUE CT FALSE & 0.23 & 0.98 & 0.05 & 0.99 & 0.05 &  & 2.24 & 0.11 & 0.08 & 0.56 & 0.05 & 0.63 \\ 
  tstats FALSE matching NA & 0.23 & 0.99 & 0.03 & 1.00 & 0.02 &  & 2.17 & 0.13 & 0.10 & 0.57 & 0.02 & 0.62 \\ 
  tstats FALSE CT FALSE & 0.23 & 0.98 & 0.06 & 0.99 & 0.05 &  & 2.24 & 0.10 & 0.07 & 0.56 & 0.05 & 0.62 \\ 
  tstats TRUE TOT NA & 0.35 & 0.89 & 0.50 & 0.95 & 0.02 &  & 2.20 & 0.12 & 0.47 & 0.59 & 0.02 & 0.63 \\ 
  CT TRUE matching NA & 0.24 & 0.97 & 0.04 & 1.00 & 0.02 &  & 2.29 & 0.08 & 0.12 & 0.55 & 0.02 & 0.35 \\ 
  CT TRUE CT FALSE & 0.24 & 0.96 & 0.06 & 0.99 & 0.06 &  & 2.35 & 0.06 & 0.16 & 0.54 & 0.06 & 0.35 \\ 
  tstats FALSE TOT NA & 0.36 & 0.90 & 0.46 & 0.95 & 0.02 &  & 2.27 & 0.10 & 0.42 & 0.57 & 0.02 & 0.62 \\ 
  CT FALSE CT FALSE & 0.24 & 0.96 & 0.07 & 0.99 & 0.06 &  & 2.34 & 0.06 & 0.12 & 0.54 & 0.06 & 0.32 \\ 
  CT FALSE matching NA & 0.24 & 0.96 & 0.06 & 1.00 & 0.02 &  & 2.34 & 0.06 & 0.13 & 0.54 & 0.02 & 0.32 \\ 
\bottomrule
\end{tabular} 
\caption{Average MSE (lower is better), proportion of correct trees (higher is better), average number of noise variables used for splitting (lower is better), pairwise prediction similarity (higher is better), and proportion of trees making the first split correctly (only for heterogeneous setting, higher is better) for the ten combinations of splitting rules and cross-validation methods included in \texttt{causalTree} that have the highest rank in terms of average MSE. Columns 2, 3, 4, 5, and 6 show simulation results when the treatment effect is homogeneous, and columns 7, 8, 9, 10, 11, and 12 show simulation results when the treatment effect is heterogeneous. Splitting rule (\texttt{split.Rule}) can be chosen from transformed outcome trees (\texttt{TOT}), causal tree (\texttt{CT}), fit-based trees (\texttt{fit}) and squared t-statistic trees (\texttt{tstats}). Adaptive (\texttt{split.Honest = FALSE}) and honest (\texttt{split.Honest = TRUE}) versions are available except for \texttt{TOT}. Cross-validation method (\texttt{cv.option}) can be chosen from transformed outcome \texttt{TOT}, matching (\texttt{matching}), \texttt{CT} and \texttt{fit}. Adaptive (\texttt{cv.Honest = FALSE}) and honest (\texttt{cv.honest = TRUE}) versions are available for \texttt{CT} and \texttt{fit}. The naming pattern for the choices are "(\texttt{split.Rule}) (\texttt{split.Honest}) (\texttt{cv.option}) (\texttt{cv.Honest})". If there is no honest version of the chosen splitting rule or cross-validation method, \texttt{split.Honest} or \texttt{cv.Honest} is \texttt{NA}.}
\label{tab: ct_settings}
\end{minipage}
}
\end{table}

\subsection{Simulations comparing the performance of regular and honest Causal Trees} \label{app: ct_nohonest}

\begin{figure}[htbp]
            \begin{center}
        \centerline{\includegraphics[width = 6in]{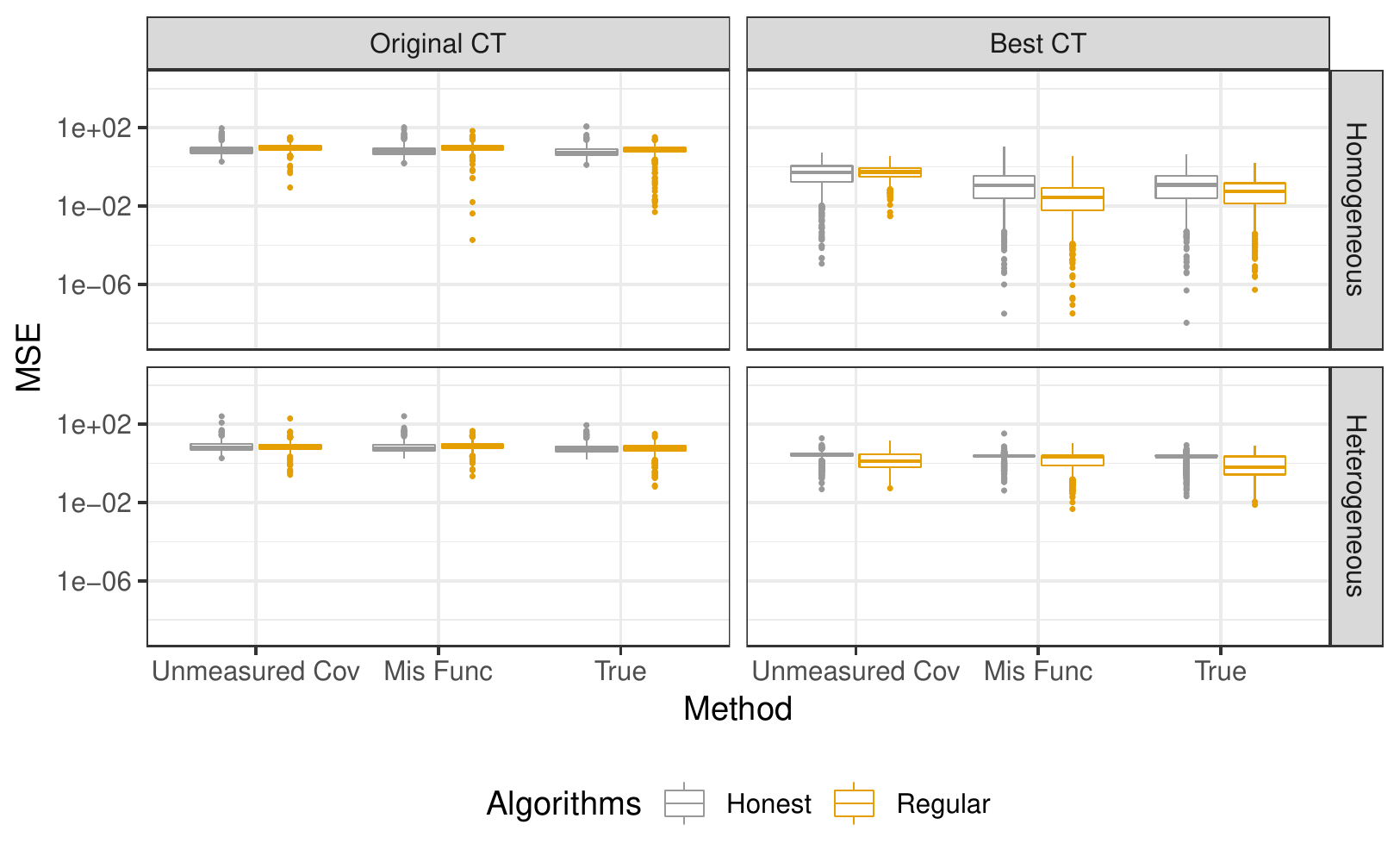}}
    \end{center}
    \caption{Mean squared error (MSE) for regular (\texttt{causalTree}) and honest (\texttt{honest.causalTree}) Causal Trees for the homogeneous (top) and the heterogeneous (bottom) settings described in Section \protect\ref{sec: sim_setup}. Lower values indicate better performance. "Original CT" refers to the original Causal Tree algorithms by \protect\cite{athey2016recursive}. "Best CT" refers to the Causal Tree algorithm with optimized splitting rule and cross-validation method. }
    \label{fig: ct_nohonest}
\end{figure}

Figure \ref{fig: ct_nohonest} presents the simulation results for the homogeneous and the heterogeneous settings described in Section \ref{sec: sim_setup} when regular (\texttt{causalTree}) and honest (\texttt{honest.causalTree}) Causal Trees are fitted using the same tuning parameter selections as for the "Original CT" and "Best CT" described in Section \ref{sec: simulations}.

Figure \ref{fig: ct_nohonest} shows that regular Causal Trees give similar or lower MSE than their honest peers for all specifications of the propensity score model. Therefore, results from regular Causal Trees are used in the comparisons presented in the main simulations in Section \ref{sec: sim-results}.

\subsection{Comparisons of running time} \label{app: cv1_time}

Table \ref{tab: time_cv1} lists the average time in seconds it takes to implement the methods compared in the simulations in Section \ref{sec: simulations}. The results show that the Doubly Robust Causal Interaction Trees (DR-CIT) run on average 8-30 times faster than Inverse Probability Weighting and G-formula Causal Interaction Trees (IPW-CIT and G-CIT). The reason is that the variance estimators for the inverse probability weighting and g-formula estimators require calculating the inverse of the quadratic form of the design matrix and thus we need to ensure the design matrix is full rank every time we calculate the splitting statistic. On the other hand, the variance estimator for the doubly robust estimator does not involve that extra step, which substantially reduces the computational complexity.

\begin{table}[htbp]
\centering
\begin{tabular}{@{}llccccccccccc@{}}
\toprule
& &Homogeneous &  & Heterogeneous\\
\cmidrule[0.5pt]{3-3} \cmidrule[0.5pt]{5-5}
Algorithm & Model & Time & & Time\\
  \hline
   Original CT &  Unmeasured Cov & 0.32 &  & 0.31 \\ 
   &  Mis Func & 0.31 &  & 0.40 \\ 
   &  True & 0.35 &  & 0.34 \\ 
   Best CT &  Unmeasured Cov & 0.06 &  & 0.06 \\ 
   &  Mis Func & 0.06 &  & 0.11 \\ 
   &  True & 0.06 &  & 0.07 \\ 
  IPW-CIT &  Unmeasured Cov & 78.14 &  & 70.55 \\ 
   &  Mis Func & 103.61 &  & 96.81 \\ 
   &  True & 80.43 &  & 74.64 \\ 
  G-CIT &  Unmeasured Cov & 229.91 &  & 226.36 \\ 
   &  Mis Func & 273.26 &  & 269.04 \\ 
   &  True & 236.79 &  & 80.65 \\ 
  DR-CIT &  Both Unmeasured Cov & 6.40 &  & 6.34 \\ 
   &  Both Mis Func & 8.20 &  & 7.07 \\ 
   &  True Prop Mis Func Out & 7.88 &  & 6.61 \\ 
   &  True Out Mis Func Prop & 9.07 &  & 8.17 \\ 
   &  Both True & 8.80 &  & 8.93 \\ 
\bottomrule
\end{tabular}
\caption{The average time in seconds it takes to implement the methods (lower is better) corresponding to algorithms in Figure \protect\ref{fig: mse} and Table \protect\ref{tab: summ_stat_cv1}. "Original CT" refers to the original Causal Tree algorithms by \protect\cite{athey2016recursive}. "Best CT" refers to the Causal Tree algorithm with optimized splitting rule and cross-validation method. IPW-CIT, G-CIT and DR-CIT refer to the Inverse Probability Weighting, G-formula, and Doubly Robust Causal Interaction Tree algorithms, respectively. "Unmeasured Cov" refers to having an unmeasured common cause of outcome and treatment assignment as described in Section \protect\ref{sec: sim_setup}. "Mis Func" refers to using a misspecified functional form of the covariates in the propensity score and/or outcome model as described in Section \protect\ref{sec: sim_setup}. "True" refers to using correctly specified propensity score and/or outcome models. For the DR estimators, "Prop" and Out" stand for propensity score model and outcome model, respectively.}
\label{tab: time_cv1}
\end{table}

\subsection{Simulations when the alternative final tree selection method described in Appendix \ref{app: FTS2} is used} \label{app: sim_fts2}

\begin{figure}[htbp]
    \centering
    \includegraphics[width = 0.9\textwidth]{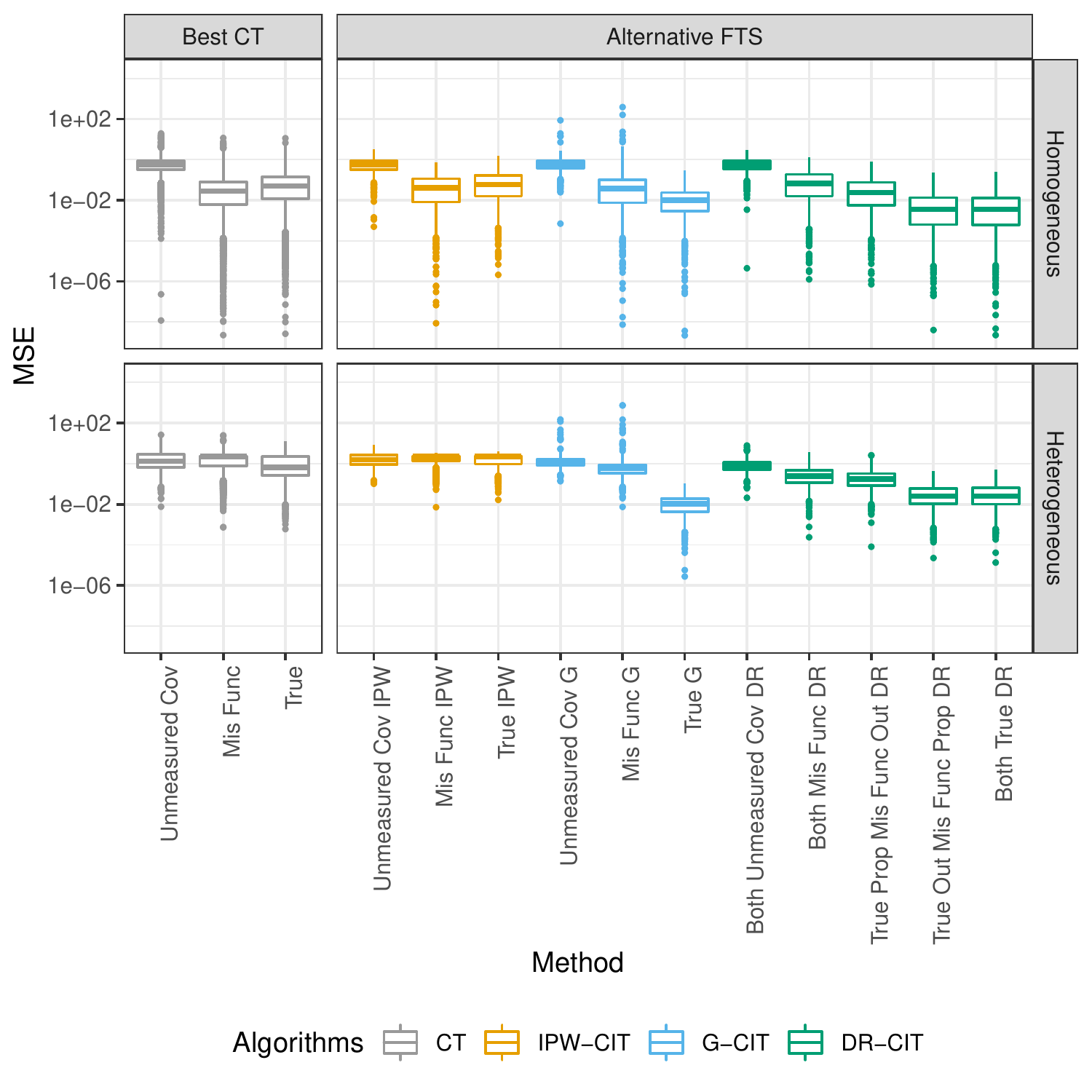}
    \caption{Mean squared error (MSE) for the different tree building algorithms for the homogeneous (top) and the heterogeneous (bottom) settings described in Section \protect\ref{sec: sim_setup} when the alternative final tree selection method described in Appendix \protect\ref{app: FTS2} is used. Lower values indicate better performance. "Best CT" refers to the Causal Tree algorithm with optimized splitting rule and cross-validation method. "Alternative FTS" refers to the Causal Interaction Tree algorithms using the final tree selection method described in Appendix \protect\ref{app: FTS2}. IPW-CIT, G-CIT and DR-CIT refer to the Inverse Probability Weighting, G-formula, and Doubly Robust Causal Interaction Tree algorithms, respectively. 
    "Unmeasured Cov" refers to having an unmeasured common cause of outcome and treatment assignment as described in Section \protect\ref{sec: sim_setup}. "Mis Func" refers to using a misspecified functional form of the covariates in the propensity score and/or outcome model as described in Section \protect\ref{sec: sim_setup}. "True" refers to using correctly specified propensity score and/or outcome models. For the DR estimators, "Prop" and Out" stands for propensity score model and outcome model, respectively.}
    \label{fig: mse_cv2}
\end{figure}

Figure \ref{fig: mse_cv2} shows boxplots of MSE and Table \ref{tab: summ_stat_cv2} shows the proportion of correct trees, average number of noise variables, pairwise prediction similarity, the average running time for both simulation settings, and the proportion of trees making a correct first split in the heterogeneous setting when the final tree selection method described in Appendix \ref{app: FTS2} is used in connection with the Causal Interaction Tree algorithms.  

\begin{table}[htbp]
\centering
\footnotesize
\begin{tabular}{@{}llccccccccccccc@{}}
\toprule
& &\multicolumn{4}{c}{Homogeneous Effect} &  & \multicolumn{5}{c}{Heterogeneous Effect}\\
\cmidrule[0.5pt]{3-6} \cmidrule[0.5pt]{8-12}
 & & Correct & Number & & & & Correct & Number & & & Correct\\
 & Model & Trees & Noise & PPS & Time & & Trees & Noise & PPS & Time & First Split \\ 
  \hline
  IPW-CIT &  Unmeasured Cov & 0.96 & 0.05 & 0.98 & 136.60 &  & 0.49 & 0.28 & 0.74 & 121.24 & 0.76 \\ 
   &  Mis Func & 0.96 & 0.05 & 0.98 & 169.56 &  & 0.15 & 0.31 & 0.60 & 159.66 & 0.21 \\ 
   &  True & 0.98 & 0.04 & 0.99 & 136.37 &  & 0.32 & 0.19 & 0.65 & 125.75 & 0.54 \\ 
  G-CIT &  Unmeasured Cov & 0.85 & 0.40 & 0.91 & 380.79 &  & 0.44 & 0.68 & 0.78 & 372.59 & 0.97 \\ 
   &  Mis Func & 0.72 & 0.92 & 0.83 & 413.32 &  & 0.48 & 0.87 & 0.79 & 409.19 & 0.97 \\ 
   &  True & 0.34 & 3.21 & 0.56 & 402.60 &  & 0.99 & 0.00 & 1.00 & 116.21 & 1.00 \\ 
  DR-CIT &  Both Unmeasured Cov & 0.98 & 0.03 & 0.99 & 27.33 &  & 0.87 & 0.08 & 0.93 & 27.45 & 0.98 \\ 
   &  Both Mis Func & 0.90 & 0.19 & 0.98 & 38.13 &  & 0.84 & 0.17 & 0.94 & 36.66 & 0.99 \\ 
   &  True Prop Mis Func Out & 0.97 & 0.06 & 0.99 & 39.28 &  & 0.92 & 0.09 & 0.96 & 37.42 & 1.00 \\ 
   &  True Out Mis Func Prop & 0.84 & 0.27 & 0.94 & 45.48 &  & 0.78 & 0.38 & 0.97 & 46.20 & 1.00 \\ 
   &  Both True & 0.84 & 0.28 & 0.94 & 44.06 &  & 0.77 & 0.39 & 0.97 & 45.02 & 1.00 \\ 
\bottomrule
\end{tabular}
\caption{Proportion of correct trees (higher is better), average number of noise variables used for splitting (lower is better), pairwise prediction similarity (higher is better), the average time in seconds it takes to implement the methods (lower is better), and proportion of trees making the first split correctly (only for heterogeneous setting, higher is better) corresponding to algorithms in Figure \ref{fig: mse_cv2} when the alternative final tree selection method described in Appendix \ref{app: FTS2} is used. Columns 3, 4, 5, and 6 show simulation results when the treatment effect is homogeneous, and columns 7, 8, 9, 10, and 11 show simulation results when the treatment effect is heterogeneous. IPW-CIT, G-CIT and DR-CIT refer to the Inverse Probability Weighting, G-formula, and Doubly Robust Causal Interaction Tree algorithms, respectively.
"Unmeasured Cov" refers to having an unmeasured common cause of outcome and treatment assignment as described in Section \protect\ref{sec: sim_setup}. "Mis Func" refers to using a misspecified functional form of the covariates in the propensity score and/or outcome model as described in Section \ref{sec: sim_setup}. "True" refers to using correctly specified propensity score and/or outcome models. For the DR estimators, "Prop" and Out" stand for propensity score model and outcome model, respectively.}
\label{tab: summ_stat_cv2}
\end{table}

Figure \ref{fig: mse_cv2} shows that the relative performance of the Causal Interaction Tree algorithms in terms of MSE is similar to what is seen in Figure \ref{fig: mse} except that in the homogeneous setting, the MSE of the inverse probability weighting trees when the functional form of the propensity score models are misspecified is slightly smaller than that when the model is correctly specified. 

The results in Figure \ref{fig: mse_cv2} and Table \ref{tab: summ_stat_cv2} show that when the final tree selection method developed in \cite{steingrimsson2019subgroup} is used, the Causal Interaction Trees outperform the Causal Trees. However, when comparing the different Causal Interaction Tree algorithms the results do not always match what is expected based on the properties of the estimators for $\mu_a(w)$ used in the tree building process. For example, the G-formula Causal Interaction Trees (G-CIT) with a correctly specified outcome model overfit (build larger trees than the true tree) in the homogeneous setting and perform worse on some evaluation measures than the G-CITs with misspecified models. Similar trend is seen for the Doubly Robust Causal Interaction Trees in both settings. 


\subsection{Simulations when the models are fitted on the whole dataset or separately within each child node} \label{app: sim_out_insplt}

\begin{figure}[htbp]
    \centering
    \includegraphics[width = \textwidth]{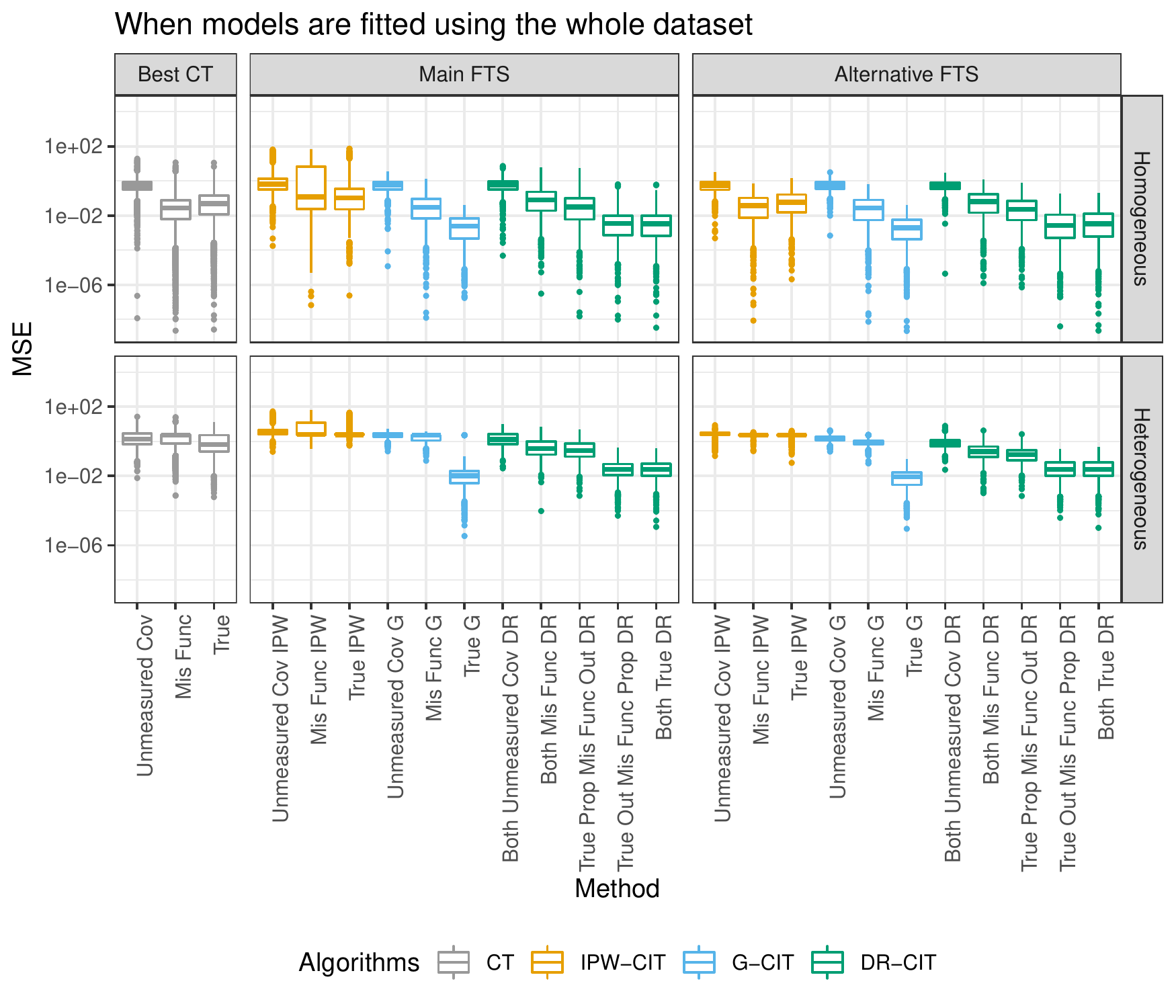}
    \caption{Mean squared error (MSE) for the different tree building algorithms for the homogeneous (top) and the heterogeneous (bottom) settings described in Section \ref{sec: sim_setup} when the propensity score and/or the outcome models are fitted prior to the tree building process using the whole dataset. Lower values indicate better performance. "Best CT" refers to the Causal Tree algorithm with optimized splitting rule and cross-validation method. "Main FTS" refers to the Causal Interaction Tree algorithms described in the main manuscript. "Alternative FTS" refers to the final tree selection method described in Appendix \ref{app: FTS2}. IPW-CIT, G-CIT and DR-CIT refer to the Inverse Probability Weighting, G-formula, and Doubly Robust Causal Interaction Tree algorithms, respectively. 
    "Unmeasured Cov" refers to having an unmeasured common cause of outcome and treatment assignment as described in Section \protect\ref{sec: sim_setup}. "Mis Func" refers to using a misspecified functional form of the covariates in the propensity score and/or outcome model as described in Section \ref{sec: sim_setup}. "True" refers to using correctly specified propensity score and/or outcome models. For DR estimators, "Prop" and Out" stands for propensity score model and outcome model, respectively.}
    \label{fig: mse_out}
\end{figure}

\begin{figure}[htbp]
    \centering
    \includegraphics[width = \textwidth]{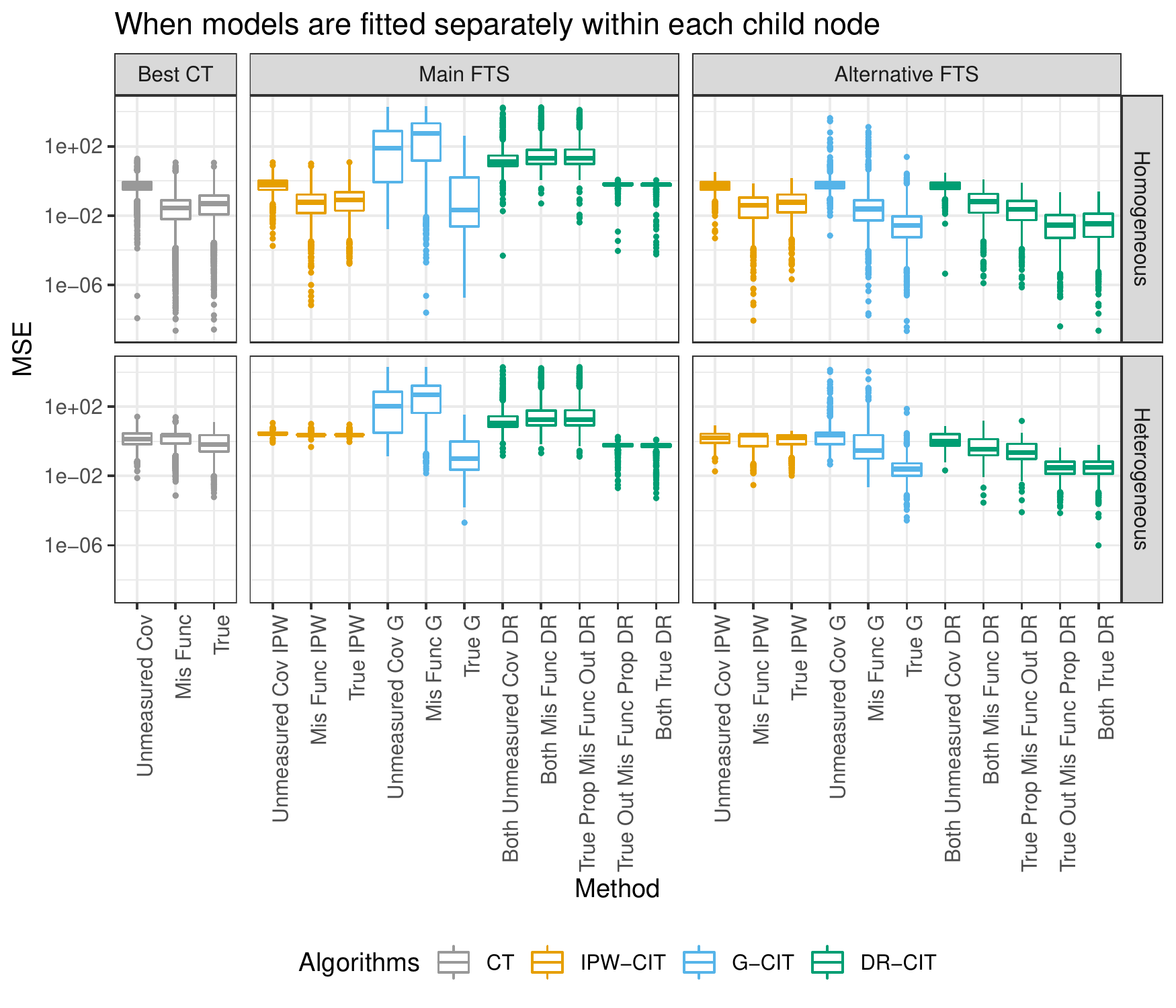}
    \caption{Mean squared error (MSE) for the different tree building algorithms for the homogeneous (top) and the heterogeneous (bottom) settings described in Section \ref{sec: sim_setup} when the propensity score and/or the outcome models are fitted within each potential child node separately. Lower values indicate better performance. "Best CT" refers to the Causal Tree algorithm with optimized splitting rule and cross-validation method. "Main FTS" refers to the Causal Interaction Tree algorithms described in the main manuscript. "Alternative FTS" refers to the final tree selection method described in Appendix \ref{app: FTS2}. IPW-CIT, G-CIT and DR-CIT refer to the Inverse Probability Weighting, G-formula, and Doubly Robust Causal Interaction Tree algorithms, respectively. "Unmeasured Cov" refers to having an unmeasured common cause of outcome and treatment assignment as described in Section \protect\ref{sec: sim_setup}. "Mis Func" refers to using a misspecified functional form of the covariates in the propensity score and/or outcome model as described in Section \ref{sec: sim_setup}. "True" refers to using correctly specified propensity score and/or outcome models. For DR estimators, "Prop" and Out" stands for propensity score model and outcome model, respectively.}
    \label{fig: mse_insplit}
\end{figure}

Figure \ref{fig: mse_out} shows boxplots of MSE and Table \ref{tab: summ_stat_out} shows the proportion of correct trees, average number of noise variables, pairwise prediction similarity, the average running time for both simulation settings, and the proportion of trees making a correct first split in the heterogeneous setting when the propensity score and/or the outcome models are fitted prior to the tree building process using the whole dataset. Figure \ref{fig: mse_insplit} and Table \ref{tab: summ_stat_insplit} show the results when the models are fitted separately in each potential child node.


\begin{table}[htbp]
\centering
\footnotesize
\begin{tabular}{@{}llccccccccccccc@{}}
\toprule
& &\multicolumn{4}{c}{Homogeneous Effect} &  & \multicolumn{5}{c}{Heterogeneous Effect}\\
\cmidrule[0.5pt]{3-6} \cmidrule[0.5pt]{8-12}
 & & Correct & Number & & & & Correct & Number & & & Correct\\
 & Model & Trees & Noise & PPS & Time & & Trees & Noise & PPS & Time & First Split \\ 
  \hline
\multicolumn{2}{@{}l}{\textbf{Main FTS}}\\
IPW-CIT & Unmeasured Cov & 0.82 & 1.77 & 0.86 & 81.32 &  & 0.04 & 1.71 & 0.55 & 74.69 & 0.71 \\
   & Mis Func & 0.69 & 2.37 & 0.80 & 129.49 &  & 0.01 & 2.53 & 0.53 & 121.13 & 0.22 \\ 
   & True & 0.85 & 1.32 & 0.90 & 82.86 &  & 0.03 & 1.32 & 0.53 & 77.50 & 0.50 \\ 
  G-CIT & Unmeasured Cov & 0.92 & 0.11 & 0.96 & 233.63 &  & 0.32 & 0.06 & 0.61 & 224.29 & 0.96 \\ 
   & Mis Func & 0.92 & 0.10 & 0.96 & 268.39 &  & 0.33 & 0.04 & 0.62 & 263.19 & 0.96 \\ 
   & True & 1.00 & 0.00 & 1.00 & 243.08 &  & 0.99 & 0.00 & 0.99 & 85.19 & 1.00 \\ 
  DR-CIT & Both Unmeasured Cov & 0.92 & 0.16 & 0.97 & 3.69 &  & 0.58 & 0.17 & 0.79 & 3.81 & 0.97 \\ 
   & Both Mis Func & 0.92 & 0.21 & 0.97 & 4.90 &  & 0.73 & 0.19 & 0.87 & 3.68 & 0.99 \\ 
   & True Prop Mis Func Out & 0.94 & 0.15 & 0.98 & 7.36 &  & 0.77 & 0.19 & 0.88 & 5.80 & 1.00 \\ 
   & True Out Mis Func Prop & 0.95 & 0.17 & 0.98 & 4.96 &  & 0.94 & 0.12 & 0.99 & 4.38 & 1.00 \\ 
   & Both True & 0.94 & 0.18 & 0.98 & 7.87 &  & 0.94 & 0.10 & 0.99 & 7.25 & 1.00 \\ 
\multicolumn{2}{@{}l}{\textbf{Alternative FTS}}\\
  IPW-CIT &  Unmeasured Cov & 1.00 & 0.00 & 1.00 & 120.36 &  & 0.18 & 0.00 & 0.57 & 110.54 & 0.76 \\ 
   &  Mis Func & 1.00 & 0.00 & 1.00 & 191.54 &  & 0.01 & 0.00 & 0.51 & 182.87 & 0.21 \\ 
   &  True & 1.00 & 0.00 & 1.00 & 125.46 &  & 0.07 & 0.00 & 0.53 & 117.80 & 0.54 \\ 
  G-CIT &  Unmeasured Cov & 0.80 & 0.82 & 0.88 & 319.02 &  & 0.59 & 1.22 & 0.78 & 303.47 & 0.97 \\ 
   &  Mis Func & 0.91 & 0.22 & 0.95 & 372.48 &  & 0.59 & 0.75 & 0.79 & 354.27 & 0.97 \\ 
   &  True & 0.23 & 20.10 & 0.48 & 323.95 &  & 0.99 & 0.00 & 1.00 & 118.23 & 1.00 \\ 
  DR-CIT &  Both Unmeasured Cov & 1.00 & 0.00 & 1.00 & 9.47 &  & 0.92 & 0.02 & 0.94 & 8.80 & 0.98 \\ 
   &  Both Mis Func & 1.00 & 0.00 & 1.00 & 12.63 &  & 0.93 & 0.02 & 0.94 & 10.70 & 0.99 \\ 
   &  True Prop Mis Func Out & 1.00 & 0.00 & 1.00 & 20.36 &  & 0.96 & 0.02 & 0.97 & 17.68 & 1.00 \\ 
   &  True Out Mis Func Prop & 0.87 & 0.20 & 0.95 & 12.26 &  & 0.77 & 0.39 & 0.97 & 11.05 & 1.00 \\ 
   &  Both True & 0.85 & 0.25 & 0.94 & 19.31 &  & 0.78 & 0.36 & 0.97 & 20.30 & 1.00 \\ 
\bottomrule
\end{tabular}
\caption{Proportion of correct trees (higher is better), average number of noise variables used for splitting (lower is better), pairwise prediction similarity (higher is better), the average time in seconds it takes to implement the methods (lower is better), and proportion of trees making the first split correctly (only for heterogeneous setting, higher is better) corresponding to algorithms in Figure \ref{fig: mse_out} when the propensity score and/or the outcome models are fitted prior to the tree building process using the whole dataset. Columns 3, 4, 5, and 6 show simulation results when the treatment effect is homogeneous, and columns 7, 8, 9, 10, and 11 show simulation results when the treatment effect is heterogeneous. "Main FTS" refers to the Causal Interaction Tree algorithms described in the main manuscript. "Alternative FTS" refers to the final tree selection method described in Appendix \ref{app: FTS2}. IPW-CIT, G-CIT and DR-CIT refer to the Inverse Probability Weighting, G-formula, and Doubly Robust Causal Interaction Tree algorithms, respectively. "Unmeasured Cov" refers to having an unmeasured common cause of outcome and treatment assignment as described in Section \protect\ref{sec: sim_setup}. "Mis Func" refers to using a misspecified functional form of the covariates in the propensity score and/or outcome model as described in Section \ref{sec: sim_setup}. "True" refers to using correctly specified propensity score and/or outcome models. For DR estimators, "Prop" and Out" stands for propensity score model and outcome model, respectively.}
\label{tab: summ_stat_out}
\end{table}

\begin{table}[htbp]
\centering
\footnotesize
\begin{tabular}{@{}llccccccccccccc@{}}
\toprule
& &\multicolumn{4}{c}{Homogeneous Effect} &  & \multicolumn{5}{c}{Heterogeneous Effect}\\
\cmidrule[0.5pt]{3-6} \cmidrule[0.5pt]{8-12}
 & & Correct & Number & & & & Correct & Number & & & Correct\\
 & Model & Trees & Noise & PPS & Time & & Trees & Noise & PPS & Time & First Split \\ 
  \hline
\multicolumn{2}{@{}l}{\textbf{Main FTS}}\\
IPW-CIT & Unmeasured Cov & 0.94 & 0.09 & 0.99 & 345.84 &  & 0.00 & 0.05 & 0.50 & 270.87 & 0.61 \\ 
   & Mis Func & 0.94 & 0.10 & 0.99 & 469.13 &  & 0.00 & 0.06 & 0.50 & 380.58 & 0.53 \\ 
   & True & 0.95 & 0.09 & 0.99 & 387.79 &  & 0.01 & 0.07 & 0.50 & 300.24 & 0.62 \\ 
   G-CIT & Unmeasured Cov & 0.32 & 10.53 & 0.57 & 1081.84 &  & 0.07 & 8.50 & 0.70 & 875.26 & 0.66 \\ 
   & Mis Func & 0.19 & 19.33 & 0.38 & 1772.99 &  & 0.05 & 14.16 & 0.69 & 1239.93 & 0.76 \\ 
   & True & 0.52 & 12.26 & 0.66 & 2106.50 &  & 0.49 & 8.01 & 0.87 & 1470.95 & 1.00 \\ 
  DR-CIT & Both Unmeasured Cov & 0.03 & 14.76 & 0.15 & 672.03 &  & 0.02 & 11.23 & 0.61 & 462.08 & 0.74 \\ 
   & Both Mis Func & 0.00 & 16.91 & 0.09 & 948.73 &  & 0.00 & 13.20 & 0.59 & 604.45 & 0.80 \\ 
   & True Prop Mis Func Out & 0.01 & 16.61 & 0.10 & 867.19 &  & 0.00 & 13.28 & 0.59 & 560.50 & 0.82 \\ 
   & True Out Mis Func Prop & 0.00 & 18.04 & 0.07 & 911.18 &  & 0.02 & 13.35 & 0.59 & 630.70 & 1.00 \\ 
   & Both True & 0.02 & 16.81 & 0.10 & 822.44 &  & 0.04 & 12.04 & 0.61 & 584.83 & 1.00 \\ 
  \multicolumn{2}{@{}l}{\textbf{Alternative FTS}}\\
  IPW-CIT &  Unmeasured Cov & 0.99 & 0.01 & 1.00 & 513.17 &  & 0.43 & 0.15 & 0.76 & 407.37 & 0.63 \\ 
   &  Mis Func & 1.00 & 0.01 & 1.00 & 714.87 &  & 0.27 & 0.21 & 0.71 & 571.29 & 0.55 \\ 
   &  True & 1.00 & 0.01 & 1.00 & 590.30 &  & 0.28 & 0.20 & 0.72 & 449.82 & 0.66 \\ 
  G-CIT &  Unmeasured Cov & 0.95 & 0.06 & 1.00 & 1576.02 &  & 0.44 & 0.23 & 0.79 & 1292.56 & 0.73 \\
   &  Mis Func & 0.94 & 0.06 & 1.00 & 2654.03 &  & 0.63 & 0.18 & 0.86 & 1852.03 & 0.83 \\ 
   &  True & 0.89 & 0.28 & 0.99 & 3067.14 &  & 0.79 & 0.49 & 0.99 & 2141.08 & 1.00 \\ 
  DR-CIT &  Both Unmeasured Cov & 1.00 & 0.00 & 1.00 & 1064.68 &  & 0.64 & 0.04 & 0.81 & 719.99 & 0.78 \\ 
   &  Both Mis Func & 0.99 & 0.01 & 1.00 & 1549.84 &  & 0.71 & 0.06 & 0.85 & 943.76 & 0.84 \\ 
   &  True Prop Mis Func Out & 1.00 & 0.00 & 1.00 & 1426.54 &  & 0.78 & 0.04 & 0.88 & 878.52 & 0.86 \\ 
   &  True Out Mis Func Prop & 0.88 & 0.21 & 0.98 & 1443.81 &  & 0.78 & 0.43 & 0.97 & 975.67 & 1.00 \\ 
   &  Both True & 0.85 & 0.26 & 0.97 & 1284.97 &  & 0.79 & 0.42 & 0.98 & 895.38 & 1.00 \\
\bottomrule
\end{tabular}
\caption{Proportion of correct trees (higher is better), average number of noise variables used for splitting (lower is better), pairwise prediction similarity (higher is better), the average time in seconds it takes to implement the methods (lower is better), and proportion of trees making the first split correctly (only for heterogeneous setting, higher is better) corresponding to algorithms in Figure \ref{fig: mse_insplit} when the propensity score and/or the outcome models are fitted within each potential child node separately. Lower values indicate better performance. Columns 3, 4, 5, and 6 show simulation results when the treatment effect is homogeneous, and columns 7, 8, 9, 10, and 11 show simulation results when the treatment effect is heterogeneous. "Main FTS" refers to the Causal Interaction Tree algorithms described in the main manuscript. "Alternative FTS" refers to the final tree selection method described in Appendix \ref{app: FTS2}. IPW-CIT, G-CIT and DR-CIT refer to the Inverse Probability Weighting, G-formula, and Doubly Robust Causal Interaction Tree algorithms, respectively. "Unmeasured Cov" refers to having an unmeasured common cause of outcome and treatment assignment as described in Section \protect\ref{sec: sim_setup}. "Mis Func" refers to using a misspecified functional form of the covariates in the propensity score and/or outcome model as described in Section \ref{sec: sim_setup}. "True" refers to using correctly specified propensity score and/or outcome models. For DR estimators, "Prop" and Out" stands for propensity score model and outcome model, respectively.}
\label{tab: summ_stat_insplit}
\end{table}

The performance of Causal Interaction Tree algorithms when the propensity score and/or the outcome models are fitted on the whole dataset presented in Figure \ref{fig: mse_out} and Table \ref{tab: summ_stat_out} is similar to that when the models are fitted using the data in the node that is being considered for splitting (Figure \ref{fig: mse}, Table \ref{tab: summ_stat_cv1}, Figure \ref{fig: mse_cv2} and Table \ref{tab: summ_stat_cv2}). 

When the models are fitted separately within each potential child node, the performance of Causal Interaction Trees using the final tree selection method described in the main manuscript becomes worse compared with those when the models are fitted using the data falling in the node being considered for splitting or using the whole dataset. In particular, the algorithms overfit to the data since the trees split on noise variables and are still able to identify the correct first split in the heterogeneous setting. A potential explanation is the instability is induced into the modeling procedure when only a small subset of the data is used to estimate the outcome and/or propensity score models.

\subsection{Simulations for a binary outcome with continuous and categorical covariates} \label{app: sim_bin_mixed}

In this section, we present simulation results where the outcome is binary and the covariate vector includes both continuous and categorical variables. The current implementation of the Causal Tree algorithms can only be implemented treating categorical covariates as ordinal.

The covariate vector included six variables. The first three components were generated from a 3-dimensional mean zero multivariate normal distribution, where $\forall j, k \in \{1, 2, 3\}$, $\text{cov}(X^{(j)}, X^{(k)}) = 0.3$ when $j \ne k$, and $\text{Var}(X^{(j)}) = 1$. For the other three components, $j\in \{4, 5, 6\}$, $X^{(j)}$ was generated from a discrete uniform distribution with $j$ levels and for convenience the levels were labelled with the first $j$ captalized letters. For example, $X^{(4)}$ took on the value of "A", "B", "C", or "D". The treatment indicator $A$ was simulated from a $\text{Bernoulli}(p)$ distribution, where $p = \text{expit} \left[ 0.3 X^{(2)} - 0.3X^{(3)} + 0.3 I(X^{(6)} \in \{\text{"B", "C"}\} ) \right]$. As in the simulations in the main manuscript, the outcome was simulated from two settings, one with a homogeneous treatment effect and one with a heterogeneous treatment effect. 
\begin{itemize}
    \item In the homogeneous treatment effect setting, the outcome $Y$ was simulated from a Bernoulli distribution with $\PP(Y=1|\bm{X}, A) = 0.15 + 0.1A +\text{expit}\left[ 0.2X^{(2)} \right] - 0.4 I(X^{(4)} \in \{\text{"B", "D"}\} ) $. For this setting, the treatment effect is the same for all covariate values and the correct tree consists only of the root node. 
    
    \item In the heterogeneous treatment effect setting, the outcome $Y$ was simulated from a Bernoulli distribution with $\PP(Y=1|\bm{X}, A) = 0.1 + 0.1A +\text{expit}\left[ 0.2X^{(2)} \right] - 0.4 A I(X^{(4)} \in \{\text{"B", "D"}\} )$. For this setting, the treatment effect differs depending on whether $X^{(4)} \in \{\text{"B", "D"}\} $ or not. The correct tree splits the dataset into $X^{(4)} \in \{\text{"B", "D"}\}$ and $X^{(4)} \in \{\text{"A", "C"}\}$ groups. 
\end{itemize}
For both simulation settings, a training and a test set were generated by drawing 1000 independent samples from the joint distribution of $(\bm{X}, A, Y)$.

The tree-based algorithms were implemented with correct model specification and two versions of misspecified models. The correct logistic regression model for estimating the propensity scores in $\hat{\mu}_{\text{IPW},a}(w)$, $\hat{\mu}_{\text{DR},a}(w)$ and the \texttt{weights} in Causal Tree algorithms includes main effects of $X^{(2)}, X^{(3)}$ and $I(X^{(6)} \in \{\text{"B", "C"}\} )$. The correct logistic regression outcome model used to implement $\hat{\mu}_{\text{G},a}(w)$, $\hat{\mu}_{\text{DR},a}(w)$ includes $A, X^{(2)}$ and $I(X^{(4)} \in \{\text{"B", "D"}\} )$ for the homogeneous setting and an interaction between $A$ and $I(X^{(4)} \in \{\text{"B", "D"}\} )$ instead of $I(X^{(4)} \in \{\text{"B", "D"}\} )$ for the heterogeneous setting. 

For the functional form model misspecification, the propensity score model includes exponentiated form of all continuous covariates and dummy coding of all categorical variables; the outcome model includes main effects of treatment and all covariates and all two way treatment-covariate interactions in the original form of all continuous variables and all dummy coded categorical variables.

We also implemented a version where there is an unmeasured covariate that is a common cause of treatment and the outcome. For that case, we exclude $X^{(2)}$ from the dataset when fitting the tree-based algorithms. The propensity score model includes main effects of all covariates except for $X^{(2)}$; the outcome model includes main effect of treatment and all covariates except for $X^{(2)}$ and all two-way treatment-covariate interactions that do not involve $X^{(2)}$.

The "Best CT" in Causal Tree algorithms in this simulation setting is when we set \texttt{split.Rule} to \texttt{"tstats"} without honest splitting (\texttt{split.Honest = FALSE}) and \texttt{cv.option} to \texttt{"fit"} without honest cross-validation (\texttt{cv.Honest = FALSE}). Additionally, regular Causal Trees (\texttt{causalTree}) give lower MSE than their honest peers (\texttt{honest.causalTree}) from simulations, so we present the results from regular Causal Trees.

Other implementation choices for Causal Tree and Causal Interaction Tree algorithms were as described in Section \ref{sec: implementation}. 

\begin{figure}[htbp]
    \centering
    \includegraphics[width = \textwidth]{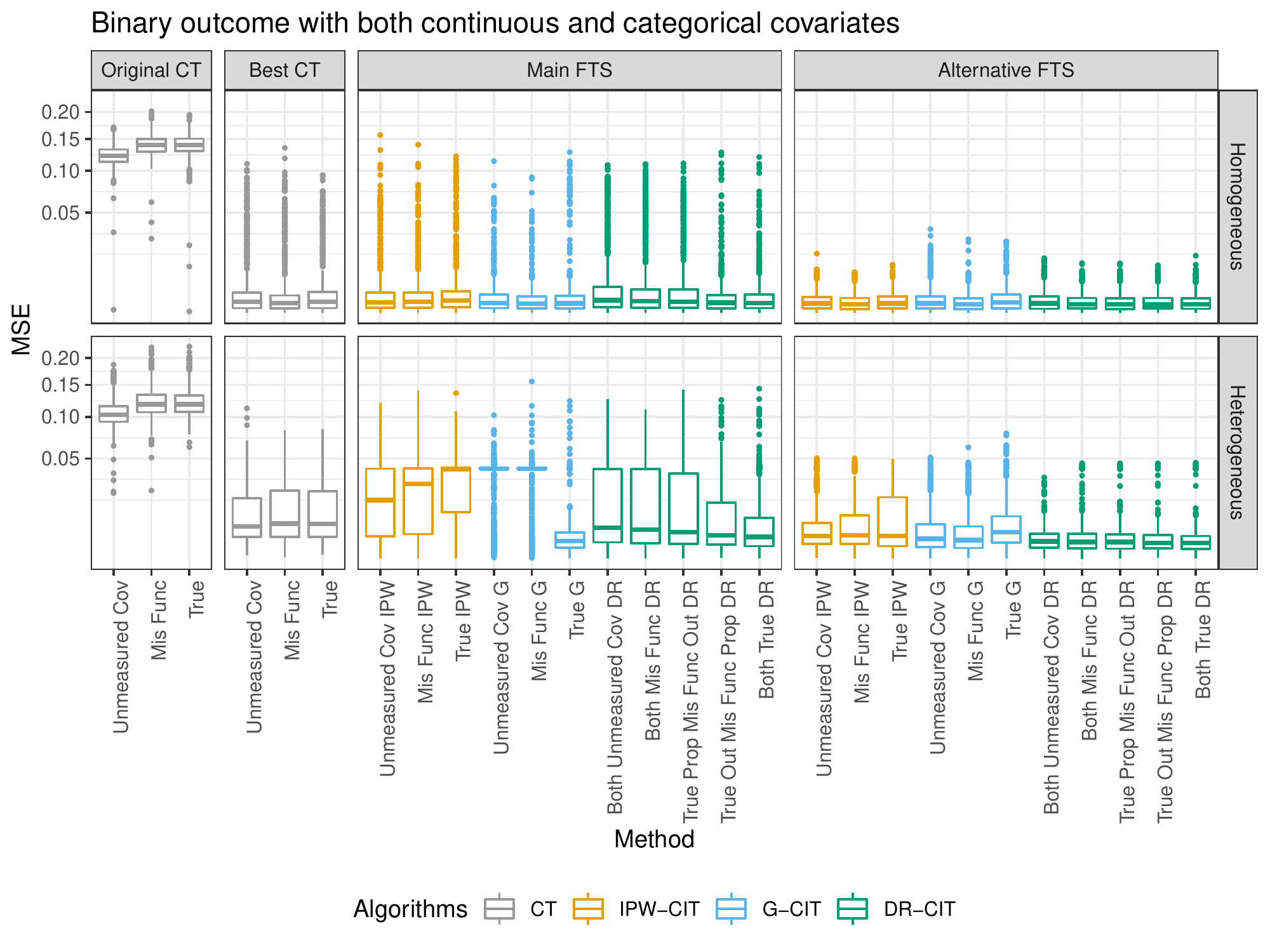}
    \caption{Mean squared error (MSE) for the different tree building algorithms for the homogeneous (top) and the heterogeneous (bottom) settings described in \ref{app: sim_bin_mixed} when the outcome is binary and the covariate vector includes both continuous and categorical variables. Lower values indicate better performance. "Original CT" refers to the original Causal Tree algorithm by \protect\cite{athey2016recursive}. "Best CT" refers to the Causal Tree algorithm with optimized splitting rule and cross-validation method. "Main FTS" refers to the Causal Interaction Tree algorithms described in the main manuscript. "Alternative FTS" refers to the final tree selection method described in Appendix \ref{app: FTS2}. IPW-CIT, G-CIT and DR-CIT refer to the Inverse Probability Weighting, G-formula, and Doubly Robust Causal Interaction Tree algorithms, respectively. "Unmeasured Cov" refers to having an unmeasured common cause of outcome and treatment assignment as described in Appendix \protect\ref{app: sim_bin_mixed}. "Mis Func" refers to using a misspecified functional form of the covariates in the propensity score and/or outcome model as described in Appendix \ref{app: sim_bin_mixed}. "True" refers to using correctly specified propensity score and/or outcome models. For DR estimators, "Prop" and Out" stands for propensity score model and outcome model respectively.}
    \label{fig: mse_bin_mixed}
\end{figure}

\begin{table}[htbp]
\centering
\footnotesize
\begin{tabular}{@{}llccccccccccccc@{}}
\toprule
& &\multicolumn{4}{c}{Homogeneous Effect} &  & \multicolumn{5}{c}{Heterogeneous Effect}\\
\cmidrule[0.5pt]{3-6} \cmidrule[0.5pt]{8-12}
 & & Correct & Number & & & & Correct & Number & & & Correct\\
 & Model & Trees & Noise & PPS & Time & & Trees & Noise & PPS & Time & First Split \\ 
  \hline
\multicolumn{2}{@{}l}{\textbf{CT}}\\
Original CT & Unmeasured Cov & 0.00 & 29.27 & 0.04 & 0.26 &  & 0.00 & 25.93 & 0.53 & 0.29 & 0.00 \\ 
   & Mis Func & 0.00 & 29.64 & 0.04 & 0.37 &  & 0.00 & 26.34 & 0.53 & 0.36 & 0.00 \\ 
   & True & 0.00 & 29.79 & 0.04 & 0.32 &  & 0.00 & 26.45 & 0.53 & 0.29 & 0.00 \\ 
  Best CT & Unmeasured Cov & 0.87 & 1.12 & 0.90 & 0.26 &  & 0.00 & 0.54 & 0.72 & 0.33 & 0.00 \\ 
   & Mis Func & 0.89 & 0.78 & 0.92 & 0.28 &  & 0.00 & 0.59 & 0.71 & 0.31 & 0.00 \\ 
   & True & 0.88 & 0.73 & 0.92 & 0.25 &  & 0.00 & 0.65 & 0.71 & 0.29 & 0.00 \\ 
\multicolumn{2}{@{}l}{\textbf{Main FTS}}\\
  IPW-CIT & Unmeasured Cov & 0.90 & 0.60 & 0.94 & 64.82 &  & 0.31 & 0.70 & 0.76 & 52.19 & 0.44 \\ 
   & Mis Func & 0.88 & 0.57 & 0.94 & 110.29 &  & 0.29 & 0.80 & 0.73 & 89.81 & 0.46 \\ 
   & True & 0.90 & 0.78 & 0.94 & 46.96 &  & 0.22 & 0.97 & 0.65 & 40.68 & 0.62 \\ 
  G-CIT & Unmeasured Cov & 0.95 & 0.49 & 0.97 & 190.79 &  & 0.15 & 0.45 & 0.60 & 174.44 & 0.90 \\ 
   & Mis Func & 0.97 & 0.29 & 0.98 & 284.57 &  & 0.16 & 0.40 & 0.59 & 253.09 & 0.91 \\ 
   & True & 0.95 & 2.45 & 0.96 & 230.68 &  & 0.95 & 0.76 & 0.98 & 253.41 & 1.00 \\ 
  DR-CIT & Both Unmeasured Cov & 0.81 & 1.83 & 0.87 & 8.15 &  & 0.58 & 2.05 & 0.86 & 8.89 & 0.98 \\
   & Both Mis Func & 0.84 & 1.35 & 0.89 & 10.35 &  & 0.59 & 1.80 & 0.86 & 10.09 & 0.97 \\ 
   & True Prop Mis Func Out & 0.84 & 1.25 & 0.90 & 11.26 &  & 0.63 & 1.62 & 0.88 & 10.36 & 0.98 \\ 
   & True Out Mis Func Prop & 0.94 & 0.28 & 0.97 & 8.44 &  & 0.72 & 0.40 & 0.90 & 9.18 & 0.97 \\ 
   & Both True & 0.93 & 0.29 & 0.97 & 9.00 &  & 0.76 & 0.39 & 0.92 & 9.46 & 0.98 \\ 
\multicolumn{2}{@{}l}{\textbf{Alternative FTS}}\\
  IPW-CIT & Unmeasured Cov & 0.98 & 0.03 & 0.99 & 104.38 &  & 0.44 & 0.14 & 0.87 & 86.97 & 0.45 \\ 
   & Mis Func & 0.98 & 0.03 & 0.99 & 173.57 &  & 0.45 & 0.16 & 0.86 & 142.68 & 0.48 \\ 
   & True & 0.99 & 0.01 & 1.00 & 79.38 &  & 0.62 & 0.26 & 0.88 & 70.38 & 0.74 \\ 
  G-CIT & Unmeasured Cov & 0.95 & 0.19 & 0.98 & 281.97 &  & 0.76 & 0.87 & 0.96 & 259.06 & 0.96 \\
   & Mis Func & 0.99 & 0.03 & 1.00 & 416.82 &  & 0.76 & 0.86 & 0.96 & 376.64 & 0.96 \\ 
   & True & 0.85 & 0.98 & 0.92 & 357.07 &  & 0.34 & 5.74 & 0.95 & 380.82 & 1.00 \\ 
  DR-CIT & Both Unmeasured Cov & 0.97 & 0.04 & 0.99 & 37.41 &  & 0.94 & 0.07 & 0.99 & 39.31 & 0.99 \\ 
   & Both Mis Func & 0.99 & 0.01 & 1.00 & 43.27 &  & 0.91 & 0.10 & 0.99 & 45.04 & 0.99 \\ 
   & True Prop Mis Func Out & 0.99 & 0.01 & 1.00 & 49.29 &  & 0.92 & 0.09 & 0.99 & 51.15 & 1.00 \\ 
   & True Out Mis Func Prop & 0.99 & 0.01 & 1.00 & 38.62 &  & 0.98 & 0.01 & 0.99 & 39.71 & 1.00 \\ 
   & Both True & 0.99 & 0.01 & 1.00 & 41.98 &  & 0.98 & 0.01 & 0.99 & 44.53 & 1.00 \\ 
\bottomrule
\end{tabular}
\caption{Proportion of correct trees (higher is better), average number of noise variables used for splitting (lower is better), pairwise prediction similarity (higher is better), the average time in seconds it takes to implement the methods (lower is better), and proportion of trees making the first split correctly (only for heterogeneous setting, higher is better) corresponding to algorithms in Figure \ref{fig: mse_bin_mixed} when the outcome is binary and the covariate vector includes both continuous and categorical variables. Columns 3, 4, 5, and 6 show simulation results when the treatment effect is homogeneous, and columns 7, 8, 9, 10, and 11 show simulation results when the treatment effect is heterogeneous. "Original CT" refers to the original Causal Tree algorithms by \protect\cite{athey2016recursive}. "Best CT" refers to the Causal Tree algorithm with optimized splitting rule and cross-validation method. "Main FTS" refers to the Causal Interaction Tree algorithms described in the main manuscript. "Alternative FTS" refers to the final tree selection method described in Appendix \ref{app: FTS2}. IPW-CIT, G-CIT and DR-CIT refer to the Inverse Probability Weighting, G-formula, and Doubly Robust Causal Interaction Tree algorithms, respectively. 
"Unmeasured Cov" refers to having an unmeasured common cause of outcome and treatment assignment as described in Appendix \protect\ref{app: sim_bin_mixed}. "Mis Func" refers to using a misspecified functional form of the covariates in the propensity score and/or outcome model as described in Appendix \ref{app: sim_bin_mixed}. "True" refers to using correctly specified propensity score and/or outcome models. For DR estimators, "Prop" and Out" stands for propensity score model and outcome model respectively.}
\label{tab: summ_stat_bin_mixed}
\end{table}

Figure \ref{fig: mse_bin_mixed} shows boxplots of MSE and Table \ref{tab: summ_stat_bin_mixed} shows the proportion of correct trees, average number of noise variables, pairwise prediction similarity, the average running time for both simulation settings, and the proportion of trees making a correct first split in the heterogeneous setting (the dataset is split into $X^{(4)} \in \{\text{"B", "D"}\}$ and $X^{(4)} \in \{\text{"A", "C"}\}$ groups) when the outcome is binary and the covariate vector includes both continuous and categorical variables. 

In general, the results follow the trends seen in the main simulations presented in Section \ref{sec: sim-results}. Additionally, in the heterogeneous setting, although Causal Trees split on fewer noise variables than some of the Causal Interaction Trees, both the proportion of correct trees and of trees making a correct first split are 0. This is because the current Causal Tree implementation give a default order to the levels in categorical variables when building trees, so the trees cannot split at points that do not preserve the order. For example, in the heterogeneous setting, the Causal Tree implementation can only split on $X^{(4)}$ based on whether or not $X^{(4)}\in \{"D"\}$, $X^{(4)}\in \{"C", "D"\}$, or $X^{(4)}\in \{"B", "C", "D"\}$ but cannot split based on whether or not $X^{(4)}\in \{"B", "D"\}$. Therefore, the trees usually first split based on whether or not $X^{(4)}\in \{"D"\}$ and produce trees with 4 terminal nodes with each level of $X^{(4)}$ in one node in our simulation setting. When the data generation procedure of the heterogeneous setting is modified so that the treatment effect differs depending on whether $X^{(4)}\in \{"C", "D"\}$, Causal Trees will identify the split point correctly.

\section{Additional Information on the Analysis of the SUPPORT Dataset} \label{app: data_analysis}

For the analysis of the SUPPORT Dataset we included the following covariates: Age, sex, race, years of education, income, type of medical insurance, primary disease category, secondary disease category, Duke Activity Status Index (DASI), do-not-resuscitate (DNR) status on day 1, cancer status, SUPPORT model estimate of the probability of surviving 2 months, APACHE score, Glasgow Coma Score, weight, temperature, mean blood pressure, respiratory rate, heart rate, PaO$_2$/FIO$_2$ ratio, PaCO$_2$, pH, white blood cell count, hematocrit, sodium, potassium, creatinine, bilirubin, albumin, 10 categories of admission diagnosis, and 12 categories of comorbidities illness. 
Additionally, following \cite{hirano2001estimation}, we a) do not use activities of daily living scale and urin output in the analysis due to large amount of missing data (75\% and 53\% missingness, respectively); and, b) create an additional indicator variable denoting if a participant's weight is recorded as 0 or not (there are 515 individuals with weight equal to 0). 

\end{document}